\documentclass[10pt, conference, letterpaper]{IEEEtran}
\IEEEoverridecommandlockouts
\usepackage{cite}
\usepackage{amsmath,amssymb,amsfonts}
\usepackage{graphicx}
\usepackage{textcomp}
\usepackage{xcolor}
\usepackage{color}
\usepackage{url}
\usepackage{bbm}
\usepackage{algorithm}
\usepackage[noend]{algorithmic}
\usepackage{bm}
 \usepackage{enumitem}
 \usepackage{amsthm}
 \usepackage{amsmath}
 \usepackage{longtable}
 \usepackage[small]{caption}
 \usepackage{tabularx}
 \usepackage{multirow}
\usepackage{eqparbox}
\usepackage{mathtools}
 \usepackage{subfigure}
\usepackage[utf8]{inputenc}

\usepackage{booktabs} 
\usepackage[caption=false,font=footnotesize]{subfig}

\newtheorem{thm}{Theorem}
\newtheorem{lem}{Lemma}

\DeclareMathOperator*{\argmax}{arg\,max}

\DeclarePairedDelimiter\ceil{\lceil}{\rceil}
\DeclarePairedDelimiter\floor{\lfloor}{\rfloor}
\def\BibTeX{{\rm B\kern-.05em{\sc i\kern-.025em b}\kern-.08em
    T\kern-.1667em\lower.7ex\hbox{E}\kern-.125emX}}
\begin{document}

\title{Decentralized Task Offloading in Edge Computing: A Multi-User Multi-Armed Bandit Approach}

\author{\IEEEauthorblockN{Xiong Wang\IEEEauthorrefmark{1},
Jiancheng Ye\IEEEauthorrefmark{2},
and John C.S. Lui\IEEEauthorrefmark{3}}
\IEEEauthorblockA{\IEEEauthorrefmark{1} National Engineering Research Center for Big Data Technology and System,\\
Services Computing Technology and System Lab, Cluster and Grid Computing Lab,\\
School of Computer Science and Technology, Huazhong University of Science and Technology, Wuhan, China}
\IEEEauthorblockA{\IEEEauthorrefmark{2} Network Technology Lab and Hong Kong Research Center, Huawei Technologies Co., Ltd, Hong Kong}
\IEEEauthorblockA{\IEEEauthorrefmark{3} Department of Computer Science and Engineering, The Chinese University of Hong Kong, Hong Kong}
\IEEEauthorblockA{E-mail: xiongwang@hust.edu.cn, yejiancheng@huawei.com, cslui@cse.cuhk.edu.hk}
\thanks{This work is supported in part by the  GRF 14200321 and CUHK:6905407. (Corresponding author: Jiancheng Ye.)}
}

\maketitle
\begin{abstract}
Mobile edge computing facilitates users to offload computation tasks to edge servers for meeting their stringent delay requirements. Previous works mainly explore task offloading when system-side information is given (e.g., server processing speed, cellular data rate), or centralized offloading under system uncertainty. But both generally fall short to handle task placement involving many coexisting users in a dynamic and uncertain environment. In this paper, we develop a \emph{multi-user} offloading framework considering \emph{unknown yet stochastic} system-side information to enable a \emph{decentralized user-initiated} service placement. Specifically, we formulate the dynamic task placement as an online multi-user multi-armed bandit process, and propose a decentralized epoch based offloading (DEBO) to optimize user rewards which are subjected under network delay. We show that DEBO can deduce the optimal user-server assignment, thereby achieving a \emph{close-to-optimal} service performance and \emph{tight} $O(\log T)$ offloading regret. Moreover, we generalize DEBO to various common scenarios such as unknown reward gap, dynamic entering or leaving of clients, and fair reward distribution, while further exploring when users’ offloaded tasks require \emph{heterogeneous} computing resources. Particularly, we accomplish a sub-linear regret for each of these instances. Real measurements based evaluations corroborate the superiority of our offloading schemes over state-of-the-art approaches in optimizing delay-sensitive rewards.
\end{abstract}

\section{Introduction}
The recent proliferation of smart devices has brought enormous popularity of many intelligent mobile applications (e.g.,  real-time face recognition, interactive gaming) which typically demand  low latency and intensive computation~\cite{satyanarayanan2017the}. Driven by emerging 5G and IoT, 90\% of the data will be generated and stored at the network edge~\cite{kelly2020internet}, making it difficult for resource-constrained mobile devices to handle such huge amount of data. To address this challenge, mobile edge computing (MEC) has emerged as a new computing paradigm  to push cloud frontier near to the network edge for supporting  computation-intensive yet delay-sensitive applications~\cite{shi2016edge}.

\begin{figure}[t]
  \centering
  \includegraphics[width=1\columnwidth]{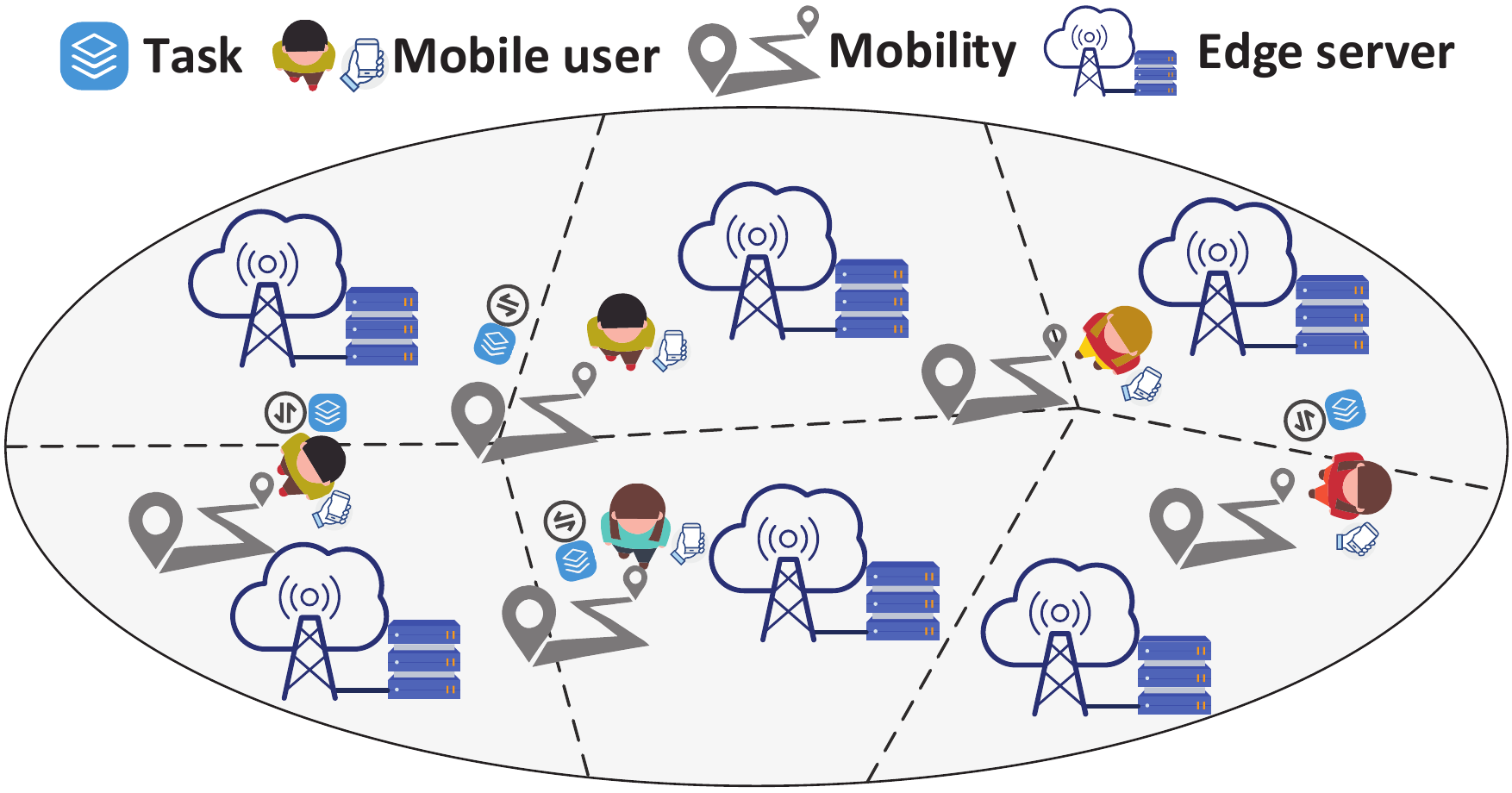}
  \caption{MEC system with separate service regions.}\label{Fig:sysmodel}
  \vspace{-10pt}
\end{figure}

With the availability of computing functionalities at the edge, MEC can facilitate mobile users to offload computation tasks to nearby edge servers, which are usually co-located with small-cell base stations and Wi-Fi access points. Under MEC,  users need to determine the service placement of their offloaded tasks so as to shorten computing latency and enhance service performance. A typical MEC system is often divided into cell regions due to  radio coverage of edge servers as shown in Fig.~\ref{Fig:sysmodel},  and this may lead to the performance disparity caused by user roaming across different service regions. Therefore, one of the core problems is to make \emph{effective offloading decision} in order to meet users' stringent delay requirements and augment servers' computing services~\cite{chen2015efficient}. 

Compared to the server-managed offloading scheme, \emph{user-initiated} task placement enables a better personalized service support tailored to their individual preference, especially when edge servers are managed by different operators~\cite{li2021online}.  However, user mobility along with the stochastic MEC environment would give rise to a time-varying service performance. Worse yet, the system-side information (e.g., server processing speed, transmission data rate, cellular  bandwidth) is usually undisclosed to mobile users, which forces the user-initiated offloading to depend on previously perceived results.

There have been various efforts devoted to task offloading in MEC systems, so as to mitigate task delay for improving computing service. Nevertheless, they generally require the complete system-side information to aid the task placement design~\cite{chen2015efficient,li2021online,sundar2018offloading}, which will be ineffective when this information is unknown or dynamically changing. Few works address system uncertainty via online learning based offloading schemes for service augmentation~\cite{sun2017EMM,ouyang2019adaptive,chen2019task}.  However, they mainly focus on \emph{centralized  offloading} with task placement decided by a \emph{single} user, while ignoring the \emph{mutual influence} of coexisting users and the \emph{capacity limitation} of edge servers.  In practice, users in MEC systems need to share the edge resources to handle computation-intensive tasks, and at the same time, are \emph{agnostic} of  the system-side information and other users' demands. A critical question we want to address is ``\emph{how to characterize the decentralized offloading for many coexisting users in an uncertain and stochastic MEC environment}''. To answer this question, researchers are faced with the following challenges.

First, unknown system-side information  demands  a \emph{learning} based adaptive offloading. In general, adaptive methods need to balance the exploration-exploitation trade-off, while the goal of achieving optimal performance for many coexisting users further complicates the offloading design. Second, a fully \emph{decentralized} placement scheme \emph{without inter-user communications} is desired as users are unaware of each other's existence when the MEC system scales. Worse still, only \emph{noisy observations} can be perceived  due to user mobility and the stochastic MEC environment, and hence the design of decentralized offloading scheme is forced to rely on bandit feedbacks. Though decentralized policies have been proposed~\cite{bistritz2018distributed,zafaruddin2019distributed,darak2019multi}, they are mainly built on the ``collision-based'' model, that is, all users would get zero rewards if making the same action, which are obviously inapplicable to model MEC systems. Third, users often have \emph{various latency-sensitivity}, which inevitably leads to distinct offloading rewards. This requires balancing their delay-sensitive rewards to ensure a fair edge resource allocation. Fourth, edge servers are endowed with \emph{limited computing capacity}, whereas different computation tasks could consume heterogeneous resource. The decentralized offloading ought to achieve a theoretically good performance while also respecting the capacity limit to avoid service blockage.

In this paper, we propose a fully decentralized  multi-user offloading scheme for MEC which does not disclose the system-side information. Considering different user  latency-sensitivity, we first devise a \emph{preference function} to characterize the offloading reward subjected under task delay, which facilitates mitigating computing latency by leveraging the \emph{perceived} reward feedback. On this basis, we formulate the dynamic task placement as  an online \emph{multi-user multi-armed bandit} (MAB) process due to the uncertain MEC environment, where offloading to an edge server is regarded as playing an arm. We develop a \emph{decentralized} epoch based offloading (DEBO) scheme to balance the offloading exploration and exploitation. As a result, the \emph{asymptotically optimal} rewards can be achieved by deriving  the optimal user-server assignment in a decentralized manner,  only using historically perceived observations. Furthermore, we advocate a heterogeneous DEBO (H-DEBO) to accommodate users' heterogeneous offloading requirements. We show that H-DEBO can ensure a good service performance even when the oracle optimal assignment is unavailable. This paper makes the following contributions.
\begin{itemize}[leftmargin=*]
\item We develop a \emph{decentralized offloading framework} for a dynamic MEC system. Our scheme achieves optimal performance for coexisting users without any inter-user communication or system-side  information. To the best of our knowledge, this is the \emph{first} work that conducts a thorough analysis of decentralized offloading under system uncertainty.
\item We propose DEBO to divide time horizon  into epochs for dynamic task placement, where each epoch consists of an exploration, matching and exploitation phase to learn the optimal user-server assignment. DEBO attains a \emph{tight} $O(\log T)$ offloading regret merely using \emph{bandit} reward feedbacks.
\item  We extend DEBO to general settings, i.e., unknown reward gap, dynamic user entering or leaving and fair reward distribution, and further quantify a \emph{sub-linear regret} for each of these extensions. More importantly, we devise the H-DEBO to accommodate \emph{heterogeneous} offloading requirements, where an $O(\log T)$ regret is accomplished by solving an APX-hard assignment problem only with \emph{learned} results. 
\item Extensive evaluations based on real measurements are performed to show the superiority of our offloading schemes over the existing approaches. In particular, we can achieve \emph{51.49\% fairness improvement} by only \emph{sacrificing 0.44\% rewards} when  incorporating the fair  reward distribution.
\end{itemize}

\section{System Model} \label{Sec: SysMdl}
We consider an MEC system with a set of mobile users $\mathcal{N}=\{1,2,...,N\}$  and edge servers $\mathcal{K}=\{1,2,...,K\}$ which can provide computing services.  When the system scales, users only retain local \emph{user indices} and need to make offloading decisions \emph{independently} as they are agnostic of each other. 

\subsection{User Task Offloading}
Each user will  offload a computation-intensive task to a server at any time $t \in \{1,2,...,T\}$ where the total time horizon $T$ is \emph{unspecified}.  Suppose $s_j$ and $f_j$ are the transmission rate and processing speed of server $j$, respectively. Also, denote $\Theta=\{(s_j,f_j), j \in \mathcal{K}\}$ as the \emph{system-side information}, which is often \emph{undisclosed} to users~\cite{ouyang2019adaptive,chen2019task}. An offloaded task from user $i$ is typically characterized by the task size $b_i$ (e.g., amount of cellular traffic)  and required CPU cycles per unit traffic $\gamma_i$~\cite{kwak2015dream}. If $i$'s task is handled by server $j$, it will experience a delay $d_{ij}$, which includes transmission and processing time:
\begin{equation}\label{Eq:delay}
d_{ij} = \frac{b_i}{s_j} + \frac{b_i\gamma_i}{f_j}. 
\end{equation}  
Along with the delay, user $i$ also associates an instant reward  $\mu_{ij}$ to represent its  personal preference on the service performance: 
\begin{equation}\label{Eq:mu}
\mu_{ij} = v_i - g_i(d_{ij}),
\end{equation}
where $v_i$ is the intrinsic task value and $g_i(\cdot)$ is a cost function which increases with delay, indicating the latency-sensitivity of user $i$. We are interested in the non-trivial case where $v_i > g_i(d_{ij})$, i.e., users acquire positive rewards after successful task offloading. Due to unknown  information $\Theta$ and system uncertainty (e.g., user mobility, transmission/processing oscillation),  each user can only perceive an i.i.d. random reward value at time $t$, with $\mu_{ij} = \mathbb{E}[r_{ij}(t)], r_{ij}(t) \in [\underline{r}, \overline{r}], \forall i \in \mathcal{N}, j \in \mathcal{K}$ where positive $\underline{r}$ and $\overline{r}$ are the lower and upper reward bounds.

\subsection{Server Computing Capacity}
Compared to the cloud datacenter, an edge server essentially has \emph{limited computing capacity}~\cite{jararweh2016the}, which we rephrase as maximum task service or  endowed computing resource depending on the user offloading requirement. 
\subsubsection{Maximum Task Service} 
When users need homogeneous resource (memory, CPU, storage) to process their offloaded tasks, the task service capacity $M_j$ of server $j$ represents how many tasks it can handle concurrently.  If excessive computation arrives,  the server will randomly choose $M_j$ tasks while abandoning the rest due to its limited capacity. Once discarded, the user experiences a high delay and observes a zero reward  $r_{ij}(t)=0$. W.l.o.g., assume  $M =\sum_{j=1}^K M_j \ge N$, i.e., MEC system has adequate resource to fulfill all offloading requests.

\subsubsection{Endowed Computing Resource} 
Considering users require heterogeneous resource for task offloading, say running different types of applications, the capacity $C_j$ of edge server $j$ stands for the amount of its endowed computing resource. 

\subsection{Problem Formulation}
\subsubsection{Homogeneous Offloading}
Under the homogeneous requirement, computing capacity of each server implies the maximum number of admitted tasks. At time $t$, denote $a_i(t) \in \mathcal{K}$  as the selected server decision of user $i$ with $\bm{a}(t) = \{a_i(t), i \in \mathcal{N}\}$, and $\mathcal{N}_j(t) = \{i, a_i(t)=j\}$ as the user set choosing server $j$. Users aim to mitigate the computing delay while avoiding task rejection due to violating the server capacity. Let $\bm{a}^*$ be the optimal offloading decisions made by an omniscient oracle when the \emph{system-side information $\Theta$ is known}, which are the solution to the offline assignment problem (OAP) below:
\begin{equation} \label{Eq:opt_num}
\begin{aligned}
\max ~&\sum_{i=1}^N  \mu_{ia_i}\\
\mathrm{s.t.}~&|\mathcal{N}_j| \le M_j, \forall j \in \mathcal{K}.
\end{aligned}
\end{equation}
Since mobile users are agnostic of each other, they can only decide $\bm{a}(t)$ in a \emph{decentralized} fashion. Accordingly, we quantify the offloading performance by its regret, defined as the difference between accumulated rewards of $\bm{a}(t)$ and that of the oracle decisions $\bm{a}^*$ to OAP:
\begin{equation} \label{Eq:regret_num}
\mathcal{R}(T) = T\sum_{i=1}^N \mu_{ia^*_i} -\sum_{t=1}^T \sum_{i=1}^N \mathbb{E}[r_{ia_i(t)}(t)].
\end{equation}
Note that $\mathbb{E}[r_{ia_i(t)}(t)] = \mu_{ia_i(t)}$ if the offloaded task is processed, otherwise  $r_{ia_i(t)}(t)=0$ once being discarded. For ease of exposition, we also refer to $\bm{a}^*$ or $\bm{a}(t)$ as an \emph{assignment} between mobile users and edge servers. 
\subsubsection{Heterogeneous Offloading}
Consider each user $i$ requires a unique $c_i$ computing resource to handle his offloaded task. Let $C'_j(t) = \sum_{i \in \mathcal{N}_j(t)} c_i$ be the total resource request for server $j$ at time $t$. Again, the optimal offloading is $\bm{a}^*$, which can be obtained by solving the heterogeneous offline assignment problem (H-OAP)  given the system-side information $\Theta$:
\begin{equation} \label{Eq:opt_heter}
\begin{aligned}
\max~&\sum_{i=1}^N  \mu_{ia_i}\\
\mathrm{s.t.}~&C'_j(t) \le C_j, \forall j \in \mathcal{K}.
\end{aligned}
\end{equation}
Different from OAP,  H-OAP is a typical generalized assignment problem (GAP) with no optimal solution for polynomial-time algorithms~\cite{cohen2006an}. Moreover,  any centralized sub-optimal solution $\bm{a}$ to GAP only ensures at most $(1+\alpha)$-approximate compound reward, that is $\sum_{i=1}^N \mu_{ia_i} \ge \frac{1}{1+\alpha} \sum_{i=1}^N \mu_{ia^*_i}, \alpha \ge 1$. As a  consequence,  the offloading regret for the heterogeneous requirement is defined by leveraging the \emph{highest guaranteed}  sub-optimal reward $\frac{1}{2}\sum_{i=1}^N \mu_{ia^*_i}$, i.e., $\alpha=1$:
\begin{equation} \label{Eq:regret_heter}
\tilde{\mathcal{R}}(T) = \frac{T}{2} \sum_{i=1}^N \mu_{ia^*_i} - \sum_{t=1}^T \sum_{i=1}^N \mathbb{E}[r_{ia_i(t)}(t)].
\end{equation}
If the task is rejected due to any server capacity limitation, we have $r_{ia_i(t)}(t) = 0$.  Also, the offloading decisions or assignment $\bm{a}(t)$ should be determined by decentralized methods. 

\subsubsection{Model Discussion} 
Our main objective is to optimize the offloading performance through dynamic task placement in a decentralized fashion. This can only be achieved by \emph{online learning} based schemes  due to the undisclosed system-side information and agnostic user coexistence. Another challenge is that because edge servers can accommodate multiple tasks, previous collision-based  indirect collaboration frameworks are  not applicable to our offloading design~\cite{bistritz2018distributed,tibrewal2019distributed}. Last but not least,  APX-hard H-OAP has no optimal solution in polynomial time caused by heterogeneous requests, which in fact demands a distinct decentralized solution  from the homogeneous OAP.


\section{Decentralized Task Offloading Scheme} \label{Sec:DEBO}
In this section, we first discuss the design of offloading scheme for the homogeneous request, with a server capacity denoting the number of admitted tasks. In particular, the user-initiated task placement is modeled as a multi-user MAB problem by regarding offloading to an edge server as playing an arm, while  it needs to be tackled via decentralized techniques.

\subsection{Epoch Based Time Division}
To achieve satisfactory service performance or minimize the offloading regret $\mathcal{R}(T)$, we have to consider the exploration-exploitation trade-off. Since the total time $T$ is unspecified aperior, we divide the time horizon into epochs  $\{1,2,...,n_T\}$, where each epoch has a variable number of  time slots and $n_T$ is the last epoch index. To strike a balance between offloading exploration and exploitation, every epoch is composed of an exploration phase, matching phase and exploitation phase.
\begin{itemize}[leftmargin=*]
\item \textbf{Exploration phase}:  this phase lasts for $T_1$ time slots in each epoch. Users will randomly offload tasks to edge servers for acquiring the estimated rewards $\tilde{\bm{r}}^{(n)} = \{\tilde{r}_{ij}^{(n)}, i \in \mathcal{N}, j\in \mathcal{K}\}$. Concretely, each $\tilde{r}_{ij}^{(n)}$ is calculated by using \emph{all explored observations} from the beginning to the current epoch $n$. 
\item \textbf{Matching phase}: this phase has a length of $T_2$ time slots. Users leverage a \emph{decentralized auction}, which will be elaborated later, based on estimated rewards $\tilde{\bm{r}}^{(n)}$ to yield an assignment $\bm{a}'$.
\item \textbf{Exploitation phase}: this phase occupies $2^n$ time slots in epoch $n$. All users offload tasks obeying  the assignment $\bm{a}'$ to fully exploit the corresponding rewards.
\end{itemize}

The fact that the exploitation takes an exponential number of  slots does not imply it occupies a long duration in practice, rather, it means the exploitation phase needs to \emph{dominate} both the exploration and matching phases. In general, our epoch division ensures the convergence of $\bm{a}'$ to the optimal assignment $\bm{a}^*$ given an accurate reward estimation $\tilde{\bm{r}}^{(n)}$.

\subsection{Decentralized Epoch Based Offloading}
\subsubsection{Algorithm Design} We now  formally specify the procedure of the decentralized epoch based offloading (DEBO) in Algorithm~\ref{Alg:DEBO}.  In epoch $n$, each user \emph{sequentially} performs the random offloading (RO) for the first $T_1$ time slots, then the decentralized auction (DAuction) for the next $T_2$ time slots, then the offloading exploitation for remaining $2^n$ time slots.


\begin{algorithm}[htp]
\begin{algorithmic}[1]
\REQUIRE{$\{M_j, j\in \mathcal{K}\}$, $M$, $T_1$, $T_2$, $\epsilon$}
\STATE  Initialization: Set  $\bm{V}=\{V_{ij} = 0,\forall i \in \mathcal{N}, j \in \mathcal{K}\}$ and $\bm{S}=\{S_{ij} = 0, \forall i \in \mathcal{N}, j \in \mathcal{K}\}$;
\FOR{epoch $n=1$ to $n_T$}
\STATE \textbf{Exploration}: $\tilde{\bm{r}}^{(n)}$= RO($\bm{V}$, $\bm{S}$, $T_1$);
\STATE \textbf{Matching}:  $\bm{a}'$= DAuction($\tilde{\bm{r}}^{(n)}$, $T_2$, $\epsilon$);
\STATE \textbf{Exploitation}: for remaining $2^n$ time slots:
\STATE~~~~User $i$ offloads  tasks to edge server $a'_i$;
\ENDFOR
\end{algorithmic}
\caption{DEBO: Decentralized Epoch Based Offloading}
\label{Alg:DEBO}
\end{algorithm}

\subsubsection{Exploration of RO} 
Users will locally execute the RO as presented in Algorithm~\ref{Alg:RO}. Since server $j$ can support $M_j$ tasks, so we regard it  has $M_j$ resource units. Accordingly, integer $H_i$ in Line 3 entails the selected server index (resource unit), and enables task abandonment if exceeding the server capacity (Line 11). Note that $\bm{V}$  and $\bm{S}$ in Line 6 record the \emph{successful} offloading times and accumulated rewards, respectively, which leverage all observations during the exploration to learn $\tilde{\bm{r}}^{(n)}$.

%

\begin{algorithm}[htp]
\begin{algorithmic}[1]
\REQUIRE{$\{M_j, j\in \mathcal{K}\}$, $M$, $\bm{V}$,  $\bm{S}$, $T_1$}
\FOR{$T_1$ time slots}
	\FOR{user $i \in \mathcal{N}$}
	\STATE Randomly choose an integer $H_i$ in $[1,M]$;
	\STATE Offload a task to the server $a_i=j$ if $H_i \in (\sum_{l=1}^{j-1}M_l, \sum_{l=1}^jM_l]$; \COMMENT{Also send integer $H_i$}
	\IF{$r_{ij}(t)>0$} 
		\STATE $V_{ij} = V_{ij} +1$,  $S_{ij} = S_{ij} + r_{ij}(t)$;
	\ENDIF
	\ENDFOR
	\FOR{edge server $j \in \mathcal{K}$}
	\IF{$|\mathcal{N}_j (t)| \le M_j$} 
		\STATE Process all offloaded tasks;
	\ELSE
		\STATE Randomly choose one task from users having the same integer $\{i, H_i=H\}$, and abandon the rest;
	\ENDIF
	\ENDFOR
\ENDFOR
\STATE  \textbf{return} $\tilde{\bm{r}}^{(n)}=\frac{\bm{S}}{\bm{V}}$;  \COMMENT{Element-wise division}
\end{algorithmic}
\caption{RO: Random Offloading}
\label{Alg:RO}
\end{algorithm}

\subsubsection{Matching of DAuction} The  DAuction  is shown in Algorithm~\ref{Alg:DAuction}. Specifically, DAuction uses the estimated rewards $\tilde{\bm{r}}^{(n)}$ to decide the assigned edge server to each user, while the optimal assignment $\bm{a}^*$ will be deduced if every learned reward $\tilde{r}_{ij}^{(n)}$ converges to the expected value $\mu_{ij}$. 

Denote range $M(m) =  (\sum_{l=1}^{j-1}M_l, \sum_{l=1}^jM_l]$  and server index  $j(m)=j$,  if $m \in (\sum_{l=1}^{j-1}M_l, \sum_{l=1}^jM_l]$. Lines 2-4 state the initialization of rewards $\bm{R}$ for each user pertaining to $M$ resource units. To simplify notation, assignment $a_i \in [1,M]$ means that user $i$ will offload tasks to server  $j(a_i)$ by holding the  resource unit $a_i$ (Line 10), while $a_i =0$ (Line 7) implies the user \emph{remains or returns} unassigned to any edge server. Based on whether we observe a positive reward (e.g., task is processed), $a_i$ is updated accordingly. In particular, resource unit $m'$ with the second highest value  in Line 9 is attained by precluding those  units belonging to the same server  to expedite the auction process~\cite{bertsekas1989the}. On edge servers' side, they will allocate their resource units (i.e., handling the offloaded tasks) to users who have the highest bids in Lines 11-13, and meanwhile discarding tasks from unassigned users. Note that the auction is decentralized as any user $i$ only needs to locally maintain a reward vector $\{R_{im}, m =1,...,M\}$, a bidding vector $\{B_{im}, m= 1,...,M\}$, and server indices $a_i, a'_i$.  
\begin{algorithm}[t]
\begin{algorithmic}[1]
\REQUIRE{$\{M_j, j\in \mathcal{K}\}$, $M$, $\tilde{\bm{r}}^{(n)}$, $T_2$, $\epsilon$}
\STATE Initialization: $\bm{R} = \{R_{im}, \forall i \in \mathcal{N}, m=1,...,M\}$, set $\bm{B} = \{B_{im}=0, \forall i \in \mathcal{N}, m=1,...,M\}$ and $\bm{a} =\bm{0}$;
\FOR{$i \in \mathcal{N}$}
	\FOR{$m=1,...,M$}
		\STATE $R_{im} = \tilde{r}_{ij}^{(n)}$ if $m \in (\sum_{l=1}^{j-1}M_l, \sum_{l=1}^jM_l]$;
	\ENDFOR	
\ENDFOR
\FOR{$T_2$ time slots}
	\FOR{user $i \in \mathcal{N}$}
	\IF{$a_i = 0$}
	\STATE Find $m^* \in \argmax_m \{R_{im}-B_{im}\}$;
	\STATE Find $m' \in \argmax_{m\notin M(m^*)} \{R_{im}-B_{im}\} $;
	\STATE Offload to server $j(m^*)$, bid $B_{im^*} = R_{im^*}-(R_{im'}-B_{im'})+\epsilon$, observe reward, set $a_i=m^*$ if task is processed and $a_i=0$ if abandoned;
	\ENDIF
	\ENDFOR
	\FOR{edge server $j \in \mathcal{K}$}
	\STATE If the $m \in (\sum_{l=1}^{j-1}M_l, \sum_{l=1}^jM_l]$-th unit is allocated to user $i$, compare all received bids with $B_{im}$ to select the highest-bid user  and allocate the $m$-th unit;
	\STATE If the $m$-th unit is not allocated, select the highest-bid  user to allocate the $m$-th unit;
	\ENDFOR
\ENDFOR
\FOR{$i \in \mathcal{N}$}
\STATE $a'_i = j(a_i)$;
\ENDFOR
\STATE  \textbf{return} $\bm{a}' = \{a'_i, i \in \mathcal{N}\}$;
\end{algorithmic}
\caption{DAuction: Decentralized Auction}
\label{Alg:DAuction}
\end{algorithm}

\subsection{Performance Analysis of DEBO}
\subsubsection{Main Regret Result}
We now analyze the regret $\mathcal{R}(T)$ in Eq.~\eqref{Eq:regret_num}. Let $\Delta_{\min} =  \min_{\bm{a} \neq \bm{a}^*} \big\{\sum_{i=1}^N  \mu_{ia^*_i} -  \sum_{i=1}^N  \mu_{ia_i}\big\}$, which is the compound reward gap between the optimal and the best sub-optimal assignments. W.l.o.g., assume there is a unique optimal assignment $\bm{a}^*$ to OAP, otherwise  $\Delta_{\min}$ implies the gap between the highest and second highest compound rewards accordingly. Moreover, denote $M_{\min} = \min_{j\in \mathcal{K}}\{M_j\}$, and $\delta_{\min} =  \min_{i \in \mathcal{N}}\min_{j,j' \in \mathcal{K},j \neq j'}\{ |\mu_{ij} -   \mu_{ij'}|\}$ as the minimum user reward gap with both $\Delta_{\min}, \delta_{\min} >0$.

\begin{thm}\label{Thm:regret_bound}
Let  $\epsilon = \max \{\frac{\Delta_{\min}}{5N}, \frac{\delta_{\min}}{K}-\frac{3\Delta_{\min}}{4NK}\}$, $T_1 =  \max \Bigl\{\ceil*{\frac{128N^2M(\overline{r}-\underline{r})^2}{9\Delta_{\min}^2 M_{\min}}}, \ceil*{\frac{81M^2}{2M_{\min}^2}}\Bigr\}$ and $T_2 = \ceil*{NM+\frac{NM \overline{r}}{\epsilon}}$. The regret $\mathcal{R}(T)$ of DEBO is upper bounded by
\begin{equation}\label{Eq:regret_bound}
\begin{aligned}
\mathcal{R}(T) &\le \Bigl(T_1N\overline{r}+T_2N\overline{r}\Bigr)\log_2(T+2) +12N^2K\overline{r}\\
&=O(\log_2T).
\end{aligned}
\end{equation}
\end{thm}
See Appendix~\ref{App:regret_bound} in for the proof. The regret bound in Eq.~\eqref{Eq:regret_bound} is obtained by \emph{bounding the error probability} $P_n$ that $\bm{a}' \neq \bm{a}^*$ after the $n$-th matching, which is shown in Lemma~\ref{Lem:EP}. Note that our derived $\log T$  regret is \emph{tight}, as a  lower $\log T$ regret bound is deduced  by regrading the user $1$ as a super-user who can control all users' offloading decisions~\cite{bistritz2020my}.

\noindent \textbf{Remark}: If  $\Delta_{\min}$ is too small, leading to a prohibitively long exploration length $T_1$, one can adjust it to a sufficiently large value in practice since $T_1$ in Theorem~\ref{Thm:regret_bound} simply provides an upper bound. Also, our regret analysis is essentially unchanged when each user $i$ can only reach a subset of servers $\mathcal{K}_i \in \mathcal{K}$.
\subsubsection{Exploration Error Probability}
The goal of exploration RO is to estimate rewards  accurately for the use in the matching phase, where the exploration error probability is presented in the next lemma with its proof in Appendix~\ref{App:EEP}. 
\begin{lem}\label{Lem:EEP}
Let $T_1 =  \max \Bigl\{\ceil*{\frac{128N^2M(\overline{r}-\underline{r})^2}{9\Delta_{\min}^2 M_{\min}}}, \ceil*{\frac{81M^2}{2M_{\min}^2}}\Bigr\}$. After the $n$-th exploration, the error probability satisfies
\begin{equation}\label{Eq:EEP}
\mathrm{Pr} \Bigl(|\tilde{r}_{ij}^{(n)} - \mu_{ij}| > \frac{3\Delta_{\min}}{8N}\Bigr) \le 3NKe^{-n}.
\end{equation}
\end{lem}
Lemma~\ref{Lem:EEP} states that each estimated reward $\tilde{r}_{ij}^{(n)}$ is sufficiently close to the expected reward $\mu_{ij}$ with high probability.

\subsubsection{Decentralized Matching Error}
Based on estimated rewards, users execute the DAuction to yield the user-server  assignment $\bm{a}'$.  If the compound reward induced by  $\bm{a}'$ is within a gap of  $\Delta_{\min}$ to the highest outcome, then  $\bm{a}'$ is in fact the optimal assignment, or $\bm{a}' =\bm{a}^*$. Note that DAuction runs in a decentralized manner with each server accommodating many tasks concurrently,  which  distinguishes it from  existing auctions  requiring inter-user communications~\cite{nayyar2018on,bertsekas1989the} or focusing on one-to-one match~\cite{naparstek2014fully, zafaruddin2019distributed}. Prior to characterizing the matching error, we  need to first show that the assignment in DAuction fulfills the $\epsilon$-complementary slackness ($\epsilon$-CS).
\begin{lem} \label{Lem:eCS}
Denote $\eta_m =\max_{i\in \mathcal{N}}\{B_{im}\}$ as the highest bid among users, and  $\tilde{\eta}_{j(m)} = \min_{m \in M(m)} \{\eta_m\}$ as the price of edge server $j(m)$ in DAuction, then the resource unit assignment $\bm{a}$ satisfies $\epsilon$-CS, that is
\begin{equation} \label{Eq:eCS}
R_{ia_i} - \tilde{\eta}_{j(a_i)} \ge \max_{m} \{R_{im}-\tilde{\eta}_{j(m)}\} - \epsilon.
\end{equation}
\end{lem}

See Appendix~\ref{App:eCS} for the proof. $\epsilon$-CS implies that the assignment $\bm{a}'$ attains at least near to the optimal reward.

\begin{lem}\label{Lem:ME}
Denote $\bm{a}^{(n)}$ as the optimal assignment under  $\tilde{\bm{r}}^{(n)}$, and $\tilde{\Delta}_{\min} = \min_{i \in \mathcal{N}, j,j' \in \mathcal{K}, j\neq j'} \{|\tilde{r}^{(n)}_{ij} - \tilde{r}^{(n)}_{ij'}|\}$. If $K\epsilon < \tilde{\Delta}_{\min}$, DAuction ensures $\bm{a}' = \bm{a}^{(n)}$. Also, there holds
\begin{equation}\label{Eq:ME}
\sum_{i=1}^N \tilde{r}^{(n)}_{ia'_i} \ge \sum_{i=1}^N \tilde{r}^{(n)}_{ia^{(n)}_i} - N\epsilon.
\end{equation}
Furthermore, DAuction will terminate with all users ended being assigned to servers within $T_2 = \ceil*{NM+\frac{NM \overline{r}}{\epsilon}}$ rounds.
\end{lem}
See Appendix~\ref{App:ME} for the proof. In addition to the commonly characterized $N\epsilon$ gap~\cite{naparstek2014fully,nayyar2018on,zafaruddin2019distributed}, our proposed DAuction further guarantees a  $K\epsilon$ matching error. By using this, we can facilitate a more precise calibration of the minimum increment $\epsilon$ in Theorem~\ref{Thm:regret_bound}. This is because the server number $K$ is usually much smaller than the user number $N$, so one can speed up DAuction by setting a larger $\epsilon$.

\subsubsection{Error Probability of DEBO}
In line with error characterizations in Lemmas~\ref{Lem:EEP} and~\ref{Lem:ME}, we present the error probability $P_n$ that $\bm{a}' \neq \bm{a}^*$ with the proof in Appendix~\ref{App:EP}.
\begin{lem} \label{Lem:EP}
If setting the parameters as in Theorem~\ref{Thm:regret_bound}, the error probability $P_n$ that $\bm{a}' \neq \bm{a}^*$ is bounded $P_n \le 3NKe^{-n}$.
\end{lem}
This lemma ensures that  $\bm{a}' $ obtained from DAuction will be optimal with increasingly high probability over epoch $n$. 

\section{Extension of Decentralized Offloading} \label{Sec:extend}
The regret outcome in Theorem~\ref{Thm:regret_bound} is materialized only when leveraging the reward gaps  $\Delta_{\min},\delta_{\min}$ and constant user number $N$. This section extends previous regret analysis to show the  robustness of our decentralized offloading in more general settings, i.e., unknown reward gaps, dynamic user entering or leaving, and fair reward distribution. 

\subsection{Unknown Reward Gap}
In previous section, we rely on the knowledge of $\Delta_{\min}$,  $\delta_{\min}$ to determine the exploration and matching lengths. In some cases, this gap information may be unavailable, thereby requiring us to set the exploration or matching length adaptively. Remember that the exploration of RO aims to acquire an accurate reward estimation, and the matching of DAuction is to deduce the optimal user-server assignment. Therefore, one can still harness DEBO in Algorithm~\ref{Alg:DEBO} by prolonging the exploration length $T_1^{(n)}$ over epoch to ensure a bounded exploration error probability, meanwhile reducing the minimum increment $\epsilon^{(n)}$, or extending the matching length $T_2^{(n)}$, so as to converge to the optimal assignment. Accordingly, we denote this decentralized offloading scheme under unknown reward gaps as U-DEBO.

\begin{thm}\label{Thm:unknownR}
Let $\epsilon^{(n)} = c_0 n^{-\vartheta}$, $T_1^{(n)} =   \ceil*{c_1 n^{\vartheta}}$ and $T_2^{(n)} = \ceil*{NM+\frac{NM \overline{r}}{\epsilon^{(n)}}}$ where $c_0$, $c_1$ are constants and $\vartheta \in (0,1)$. The regret $\mathcal{R}(T)$ of U-DEBO is bounded by
\begin{equation}\label{Eq:unknownR}
\begin{aligned}
\mathcal{R}(T) &\le \ceil*{c_1 \log_2^{\vartheta}(T+2)}N\overline{r} \log_2(T+2) \\
 &+  \ceil*{MN+\log_2^{\vartheta}(T+2) MN\overline{r}/c_0} N\overline{r} \log_2(T+2) \\
 & + N\overline{r} (2^{n_0}-1) + 12N^2K\\
 &= O\left( \log_2^{1+\vartheta} T\right),
\end{aligned}
\end{equation}
where $n_0$ is a finite integer.
\end{thm}
Please refer to Appendix~\ref{App:unknownR} for the proof. Note that there is a power $(1+\vartheta)$ on $\log T$, which stems from cautiously elongating $T_1^{(n)}$ and $T_2^{(n)}$ to ensure a bounded error probability analogous to that in Lemma~\ref{Lem:EP}.
\subsection{Dynamic User Mobility}
Mobile users may enter or leave the MEC system dynamically, leading to a time-varying user population $N$.  When this happens, the user will  notify its nearby edge server, so that the remaining users  could perceive the dynamic value of $N$ from servers without  inter-user communications. Sharing a similar spirit to~\cite{darak2019multi}, we postulate that signaling  the entering or leaving occurs at the \emph{beginning} of each epoch.

Denote $N^{(n)}$  as the user number in epoch $n$ with  $N^{(n)} \le M$ for system stability. Note that user entering and leaving will have different impacts. For the leaving case, even all users adhere to the DEBO with unchanged parameters $T_1$ and $T_2$ as in Theorem~\ref{Thm:regret_bound}, it still yields an $O(\log_2 T)$ regret.  This is because the error probability satisfies $P_n \le 3N^{(n)}Ke^{-n}$ when running RO and DAuction for longer time due to the reduction of user number $N^{(n)}$. Nevertheless, adjusting $T_1$ and $T_2$ along with $N^{(n)}$ can in fact reduce the empirical regret.  In contrast, dynamic entering causes an increase in $N^{(n)}$, which brings about an unbounded error probability  $P_n$ as newly joined users have not yet experienced adequate explorations to acquire an accurate reward estimation for attaining the optimal assignment in the matching phase. Similarly, refer to the decentralized offloading for dynamic entering or leaving as D-DEBO, whose regret is quantified  below.
\begin{thm}\label{Thm:DMEC}
Suppose the epoch $n'$ of the last user entering  satisfies $n' \le O(\log_2 T^{\zeta}), \zeta \in (0,1)$. Let $\epsilon = \max \{\frac{\Delta_{\min}}{5N^{(n)}}, \frac{\delta_{\min}}{K}-\frac{3\Delta_{\min}}{4N^{(n)}K}\}$, $T_1 =  \max \Bigl\{\ceil*{\frac{128\left(N^{(n)}\right)^2M(\overline{r}-\underline{r})^2}{9\Delta_{\min}^2 M_{\min}}}, \ceil*{\frac{81M^2}{2M_{\min}^2}}\Bigr\}$ and $T_2 = \ceil*{N^{(n)}M+\frac{N^{(n)}M \overline{r}}{\epsilon}}$. The regret $\mathcal{R}(T)$ of  D-DEBO is bounded $\mathcal{R}(T) \le O(T^{\zeta})$.
\end{thm}
See Appendix~\ref{App:DMEC} for the proof. The underlying reason for restricting the last entering time is to impede a prohibitively large \emph{exploitation regret} caused by the  error-prone reward estimation of newly joined users. 

\subsection{Fair Reward Distribution}
So far, we have focused on the utilitarian compound reward optimization, which may cause \emph{unfair server resource allocation}. Because users with higher rewards  are more likely to be assigned to servers with faster transmission rate and processing speed, thereby blocking other users from the ``better'' edge servers. To address this issue, we explore the proportional fairness maximization to enable a fair reward distribution among users, by changing the objective of OAP in Eq.~\eqref{Eq:opt_num} to $\max  \sum_{i=1}^N \ln (1+ \beta \mu_{ia_i}), \beta > 0$~\cite{kelly1997charging,wang2018dynamic}. Accordingly, let $\bm{a}^{f}$ be the optimal assignment to the egalitarian fairness optimization, and define the fairness regret as $\mathcal{R}^f(T) = T \sum_{i=1}^N \ln \big(1+ \beta \mu_{ia^f_i}\big) - \sum_{t=1}^T \sum_{i=1}^N \ln \left(1+ \beta \mathbb{E}[r_{ia_i(t)}(t)]\right)$.  

We can still employ the DEBO in Algorithm~\ref{Alg:DEBO}  by replacing the reward estimations with their logarithm fairness values. Nevertheless, this operation is built on the premise that the fairness $\ln \big(1+ \beta \tilde{r}_{ij}^{(n)} \big)$ is sufficiently accurate if the estimated reward $\tilde{r}_{ij}^{(n)}$ converges to the expected reward $\mu_{ij}$. 
\begin{lem} \label{Lem:logE}
Suppose the estimation error in RO Algorithm~\ref{Alg:RO} satisfies $|\tilde{r}_{ij}^{(n)} - \mu_{ij}| \le \Delta, \forall i \in \mathcal{N}, j \in \mathcal{K}$ with $\Delta < \frac{1+\beta \underline{r}}{4 \beta}$, then the fairness error fulfills
\begin{equation} \label{Eq:logE}
\big|\ln \big(1+ \beta \tilde{r}_{ij}^{(n)} \big)- \ln (1+ \beta \mu_{ij}) \big| \le \frac {4\beta \Delta}{3(1+\beta \underline{r})}.
\end{equation}
\end{lem}
See Appendix~\ref{App:logE} for the proof.  Based on this lemma, we adapt the DEBO to a fair decentralized offloading F-DEBO, where  the input to DAuction is $\ln \big(1+ \beta \tilde{r}_{ij}^{(n)} \big)$. Also, denote $\Delta_{\min}^f =  \min_{\bm{a} \neq \bm{a}^f}\big\{\sum_{i=1}^N \ln \big(1+ \beta \mu_{ia^f_i}\big)  -  \sum_{i=1}^N  \ln (1+ \beta \mu_{ia_i}) \big\}$ and  $\delta_{\min}^f =  \min_{i \in \mathcal{N}}\min_{j,j' \in \mathcal{K},j \neq j'}\{ |\ln (1+ \beta \mu_{ij})  - \ln (1+ \beta \mu_{ij'})|\}$. Theorem~\ref{Thm:freget} presents the fairness regret, while the proof is omitted since  it is similar to Theorem~\ref{Thm:regret_bound}. 
\begin{thm}\label{Thm:freget}
Let $\epsilon = \max \big\{\frac{\Delta_{\min}^f}{5N}, \frac{\delta_{\min}^f}{K}-\frac{3\Delta_{\min}^f}{4NK}\big\}$, $T_1 =  \max \Bigl\{\ceil*{\frac{2048N^2M(\overline{r}-\underline{r})^2 \beta^2}{81(\Delta_{\min}^f)^2 M_{\min}(1+\beta \underline{r})^2}}, \ceil*{\frac{8M(\overline{r}-\underline{r})^2 \beta^2}{M_{\min}(1+\beta \underline{r})^2}},\ceil*{\frac{81M^2}{2M_{\min}^2}}\Bigr\}$ and $T_2 = \ceil*{NM+\frac{NM \ln (1+\beta \overline{r})}{\epsilon}}$. The fairness regret $\mathcal{R}^f(T)$ of  F-DEBO  is bounded by
\begin{equation}
\begin{aligned}
\mathcal{R}^f(T) & \le T_1N \ln (1+\beta \overline{r}) \log_2(T+2) \\
& +T_2N \ln (1+\beta \overline{r}) \log_2(T+2)\\
&+12N^2K \ln (1+\beta \overline{r})\\
& = O(\log_2 T).
\end{aligned}
\end{equation}
\end{thm}

\section{Heterogeneous Decentralized Offloading}\label{Sec:HDEBO}
This section explores the task offloading under the heterogeneous resource requirement, where the server capacity stands for its endowed computing resource. Different from OAP, H-OAP is an APX-hard GAP problem with no optimal solution in polynomial time~\cite{cohen2006an}. We therefore develop a new decentralized learning to handle users' heterogeneous requests.

\subsection{Offline Problem Revisit}
In contrast to OAP, one can only attain  at most $(1+\alpha)$-approximate assignment $\bm{a}$ to H-OAP, i.e., $\sum_{i=1}^N \mu_{ia_i} \ge \frac{1}{1+\alpha} \sum_{i=1}^N \mu_{ia^*_i}$, where $\alpha \ge 1$ is the approximation ratio to the following knapsack sub-problem:
\begin{equation} \label{Eq:knapsack}
\begin{aligned}
\max~&\sum_{i \in \mathcal{N}_j} \mu_{ij}\\
\mathrm{s.t.}~ &C'_j \le C_j.
\end{aligned}
\end{equation}
In fact, $\alpha = 1$ implies deriving the optimal solution to Eq.~\eqref{Eq:knapsack}, which is also the underlying reason why we set the benchmark for the regret definition in Eq.~\eqref{Eq:regret_heter} as $\frac{1}{2}\sum_{i=1}^N \mu_{ia_i^*}$.

To obtain a $2$-approximate assignment, we first decouple the knapsack sub-problems, then acquire the optimal solution to each sub-problem sequentially. For this purpose, we introduce an 
 indicator $\bm{I}=\{I_i, i \in \mathcal{N}\}$ to record the \emph{current assignments} of all users, say $I_i = j$ means user $i$ will offload tasks to server $j$, so as to enable the decoupling of  $K$ sub-problems. Specifically,  initialize each user as unassigned $\bm{I}=\bm{0}$, and define  the reward  $\Delta \mu_{ij}$ to mark the performance improvement if user $i$'s offloaded task is processed by another edge server:
\begin{equation} \label{Eq:treward}
\Delta \mu_{ij}=
\left \{
\begin{aligned}
&\mu_{ij}  ~&& \mathrm{if}~I_i=0,\\
&\mu_{ij}-\mu_{ij'}~&&\mathrm{if}~I_i=j'.
\end{aligned}
\right.
\end{equation}
According to~\cite{cohen2006an}, we solve Eq.~\eqref{Eq:knapsack}  corresponding to each server $j$ by using $\Delta \mu_{ij}, i \in \mathcal{N}$, i.e., gradually improve the offloading reward when allocating servers' computing resource. If the solution to Eq.~\eqref{Eq:knapsack} results in user $i$ being assigned to server $j$, then update $I_i=j$ and recalculate $\Delta \mu_{ij}$ when dealing with the next sub-problem for server $j \rightarrow j+1$. As a result, the final indicator $\bm{I}$  is guaranteed to be a  $2$-approximate assignment to  H-OAP if one can \emph{optimally tackle all} $K$ sub-problems. Considering the knapsack sub-problem is NP-hard, we implement a branch-and-bound method to efficiently search its optimal solution for ensuring $\alpha=1$~\cite{neapolitan2004foundations}.

\subsection{Heterogeneous Decentralized Epoch Based Offloading}
\subsubsection{Algorithm Design}
The analysis of solving H-OAP provides us a guide for designing dynamic offloading under the  heterogeneous requirement. Similar to DEBO using learned rewards to attain the user-server assignment because of the unknown  system-side information $\Theta$, we also utilize the \emph{reward estimation} $\tilde{\bm{r}}^{(n)}$ to deduce this assignment following the characterized indicator decoupling approach. Concretely, we show the heterogeneous decentralized epoch based offloading (H-DEBO) in Algorithm~\ref{Alg:H-DEBO}, with each epoch also consisting of an exploration phase, matching phase and exploitation phase. Denote $\overline{M}_{\min} = \frac{\min_{j\in \mathcal{K}}\{C_j\}}{\max_{i\in \mathcal{N}}\{c_i\}}$ as the ratio between the minimum server capacity and maximum user requirement.

\begin{algorithm}[htp]
\begin{algorithmic}[1]
\REQUIRE{$\{C_j, j\in \mathcal{K}\}$, $\{c_i, i\in \mathcal{N}\}$, $\overline{M}_{\min}$, $K$, $T_1$}
\STATE   Initialization: Set estimated rewards  $\tilde{\bm{r}}^{(n)}=\{\tilde{r}_{ij}^{(n)}= 0,\forall i \in \mathcal{N}, j \in \mathcal{K}\}$;
\FOR{epoch $n=1$ to $n_T$}
	\STATE Users form $N_g$ groups with group size being $\overline{M}_{\min}$; \COMMENT{\textbf{Start Exploration Phase}}
	\FOR{$t=1$ to $T_1$}
	\IF{$N_g \le K$}
		\STATE Users in group $k$ offload tasks to server $[(t-1)\%K + k]\%(K+1)$,  update $\tilde{r}_{ij}^{(n)}$ using $r_{ij}(t)$;
	\ELSE
		\FOR{$g=1$ to $\floor*{N_g/K}$}
		\STATE Users in group $k, k = (g-1) K+1,...,gK$ offload tasks to server $[(t-1)\%K + k-(g-1)K]\%(K+1)$,  update $\tilde{r}_{ij}^{(n)}$ using $r_{ij}(t)$;
		\ENDFOR
		\STATE Users in group $k, k = \floor*{N_g/K} K+1,...,N_g$ offload tasks to server $[(t-1)\%K + k-\floor*{N_g/K}K]\%(K+1)$,  update $\tilde{r}_{ij}^{(n)}$ using $r_{ij}(t)$;
	\ENDIF
	\ENDFOR
	\STATE Initialize the indicator vector $\bm{I} =\{I_i  = 0, i \in \mathcal{N}\}$;  \COMMENT{\textbf{Start Matching Phase}}
	\FOR{edge server $j \in \mathcal{K}$}
	\STATE Each user $i$ computes $\Delta\tilde{r}_{ij}^{(n)}$ similar to Eq.~\eqref{Eq:treward}, offloads a task and sends $\Delta\tilde{r}_{ij}^{(n)}$ to server $j$;
	\STATE Server $j$ performs branch-and-bound to solve sub-problem of Eq.~\eqref{Eq:knapsack} with input $\{\Delta\tilde{r}_{ij}^{(n)}, i \in \mathcal{N}\}$;
	\STATE Each user $i$ updates $I_i = j$ if assigned to server $j$;
	\ENDFOR
	\STATE Set $\bm{a}' = \bm{I}$; \COMMENT{Assignment from matching}
	\FOR{remaining $2^n$ time slots}
	\STATE Each user $i$ offloads  tasks to the assigned server $a'_i$;  \COMMENT{\textbf{Exploitation Phase}}
	\ENDFOR
\ENDFOR
\end{algorithmic}
\caption{H-DEBO:  Heterogeneous Decentralized Epoch Based Offloading}
\label{Alg:H-DEBO}
\end{algorithm}

\subsubsection{Exploration Phase}
Due to users' heterogeneous requests, random offloading by holding a specific resource unit in RO  (Line 3 in Algorithm~\ref{Alg:RO}) is no longer applicable, since it is based on equally splitting the server capacity into multiple units. Therefore, we propose  a \emph{group offloading} scheme in a \emph{round-robin} fashion for reward exploration, where the group size is set to $\overline{M}_{\min}$ so not to violate any server capacity. Lines 5-6 are for group number $N_g \le K$ while Lines 8-10 amount for $N_g > K$. The exploration phase lasts for exactly $T_1$ time slots, i.e., $T_1$ observations with at least $\floor*{T_1/K}$ samples from each server for reward estimation $\bm{r}^{(n)}$.  It should be noted that the group offloading remains decentralized because any user $i$ can determine its  group only based on the local user index $i$.

\subsubsection{Matching and Exploitation Phases}
Similar to previous analysis, the matching phase mainly entails solving the knapsack sub-problem of Eq.~\eqref{Eq:knapsack} successively. To this end,  users will locally compute the rewards $\Delta \tilde{r}_{ij}^{(n)}$, which are sent to each server for updating the indicator $\bm{I}$ (i.e., offloaded task is processed) based on the  branch-and-bound method (Lines 12-15). If  every estimated reward $\tilde{r}_{ij}^{(n)}$ closely approaches the expected reward $\mu_{ij}$, the obtained $\bm{a}'$ is guaranteed to be a $2$-approximate assignment. Also, the matching phase lasts for $K$ time slots corresponding to $K$ edge  servers. Finally, users will exploit the assignment $\bm{a}'$ for $2^n$ time slots, while other dominated exploitation length enables the same effect.

\subsection{Performance Analysis of H-DEBO}
Now we analyze  the regret $\tilde{\mathcal{R}}(T)$  in Eq.~\eqref{Eq:regret_heter}. Let $\delta^{(1)}_{\min} = \min_{i,i'\in \mathcal{N}, i\neq i'} \min_{j,k \in \mathcal{K}, j \neq k}|\mu_{ij} - (\mu_{i'j}-\mu_{i'k})|$, $\delta_{\min}^{(2)} = \min_{i,i'\in \mathcal{N}, i\neq i'} \min_{j,j',k \in \mathcal{K}, j \neq j', j\neq k}|(\mu_{ij}-\mu_{ij'}) - (\mu_{i'j}-\mu_{i'k})|$, $\delta_{\min} =  \min_{i \in \mathcal{N}}\min_{j,j' \in \mathcal{K},j \neq j'}\{ |\mu_{ij} -   \mu_{ij'}|\}$. Denote $\delta'_{\min} = \min \big\{\delta^{(1)}_{\min},\delta^{(2)}_{\min},\delta_{\min}\big\}$, $\overline{M}_{\max} =\frac{\max_{j\in \mathcal{K}} \{C_j\}}{\min_{i \in \mathcal{N}}\{c_i\}}$. The following theorem states the regret with its proof in Appendix~\ref{App:heter_regret}.
\begin{thm}\label{Thm:heter_regret}
Let  $T_1 =  \ceil*{\frac{25 \overline{M}_{\max}^2(\overline{r}-\underline{r})^2}{2 (\delta'_{\min})^2}K}$ if $N_g \le K$ and $T_1 =  \ceil*{\frac{25 \overline{M}_{\max}^2(\overline{r}-\underline{r})^2}{2 (\delta'_{\min})^2}(K+N)}$ if $N_g > K$. The regret $\tilde{\mathcal{R}}(T)$ of H-DEBO is upper bounded by
\begin{equation}\label{Eq:hregret_bound}
\begin{aligned}
\tilde{\mathcal{R}}(T) &\le \Bigl(T_1\frac{N\overline{r}}{2}+K\frac{N\overline{r}}{2}\Bigr)\log_2(T+2) +4N^2K\overline{r}\\
&=O(\log_2T).
\end{aligned}
\end{equation}
\end{thm}
\noindent \textbf{Remark}: All extensions in Section ~\ref{Sec:extend} can be applied to H-DEBO, which are omitted due to page limit. Akin to DEBO, we could adjust $\delta'_{\min}$ to a larger value in practice to prevent too large  $T_1$. Besides, the accumulated rewards may exceed $\frac{T}{2} \sum_{i=1}^N \mu_{ia^*_i}$ which is in fact a theoretically lower bound.

\section{Performance Evaluation}\label{Sec:performance}
In this section, we show the evaluated results of our proposed decentralized offloading schemes and baseline methods.

\subsection{Evaluation Setup}
\subsubsection{Parameter Setting}
Consider the MEC system (divided into cells) is composed of $K=3$ edge servers and $N=8$ to $N=12$ mobile users. In line with the real measurements~\cite{kwak2015dream}, task size $b_i$ is distributed in $[500,1600]$ KB with required CPU cycles  per bit being $\gamma_i = 1000$. Server processing speed $f_j$ is in $[4,8]$ GHz, and cellular transmission rate $s_j$ is set in  $[9,11]$ Mbps  according to  the typical 4G uplink speed~\cite{4G4U}. Server capacity for maximum task service $M_j$ is an integer in $[2,5]$, while for  endowed computing resource $C_j$ is randomly in $[2,3.5]$ with user resource requirement $c_i \in [0.5,1]$. The reward preference function of Eq.~\eqref{Eq:mu} is $\mu_{ij}= v_i -\rho_i d_{ij}$ with task value $v_i \in [3,3.5]$ and $\rho_i \in [0.2,0.5]$. The random reward observation $r_{ij}(t) \in [\mu_{ij}-0.3, \mu_{ij}+0.3]$, and also let $\underline{r}=0.3, \overline{r}=3.8$ accordingly. The total number of time slots is $T=6\times10^6$. 

\subsubsection{Benchmark Algorithms}
To show the effectiveness of our proposed DEBO, its extensions and H-DEBO, we introduce the following algorithms as benchmarks for comparison.
\begin{itemize}[leftmargin=*]
\item M-UCB: upper confidence bound (UCB) algorithm which  pulls arm with the highest confidence bound of the average reward~\cite{auer2002finite}. Considering that there are multiple users, suppose they independently execute UCB for task offloading, namely M-UCB.
\item M-EXP3:  exponential-weight algorithm for exploration and exploitation (EXP3)  that pulls arms following the exponential-weight probability~\cite{auer2002the}. For our problem, multiple users will locally choose edge servers based on EXP3, i.e., M-EXP3.
\item  DM-Non0: decentralized multi-user MAB  with non-zero rewards~\cite{magesh2019decentralized}. To the best of our knowledge, this is the only work that studies non-collision model, where the reward of any individual user also depends on the user number playing the same arm.  
\end{itemize} 
The optimal assignment $\bm{a}^*$ to OAP  is obtained through  Hungarian algorithm~\cite{kuhn1955the} by splitting each server $j$ into $M_j$ copies, and is utilized to compute the regret $\mathcal{R}(T)$ in Eq.~\eqref{Eq:regret_num}.


\subsection{Evaluation Results over Time}
We first show the performance over time given $N=10$.
\begin{figure}[t]
  \centering
   \subfigure[Accumulated rewards]{\includegraphics[width=0.49\linewidth,height=1.3in]{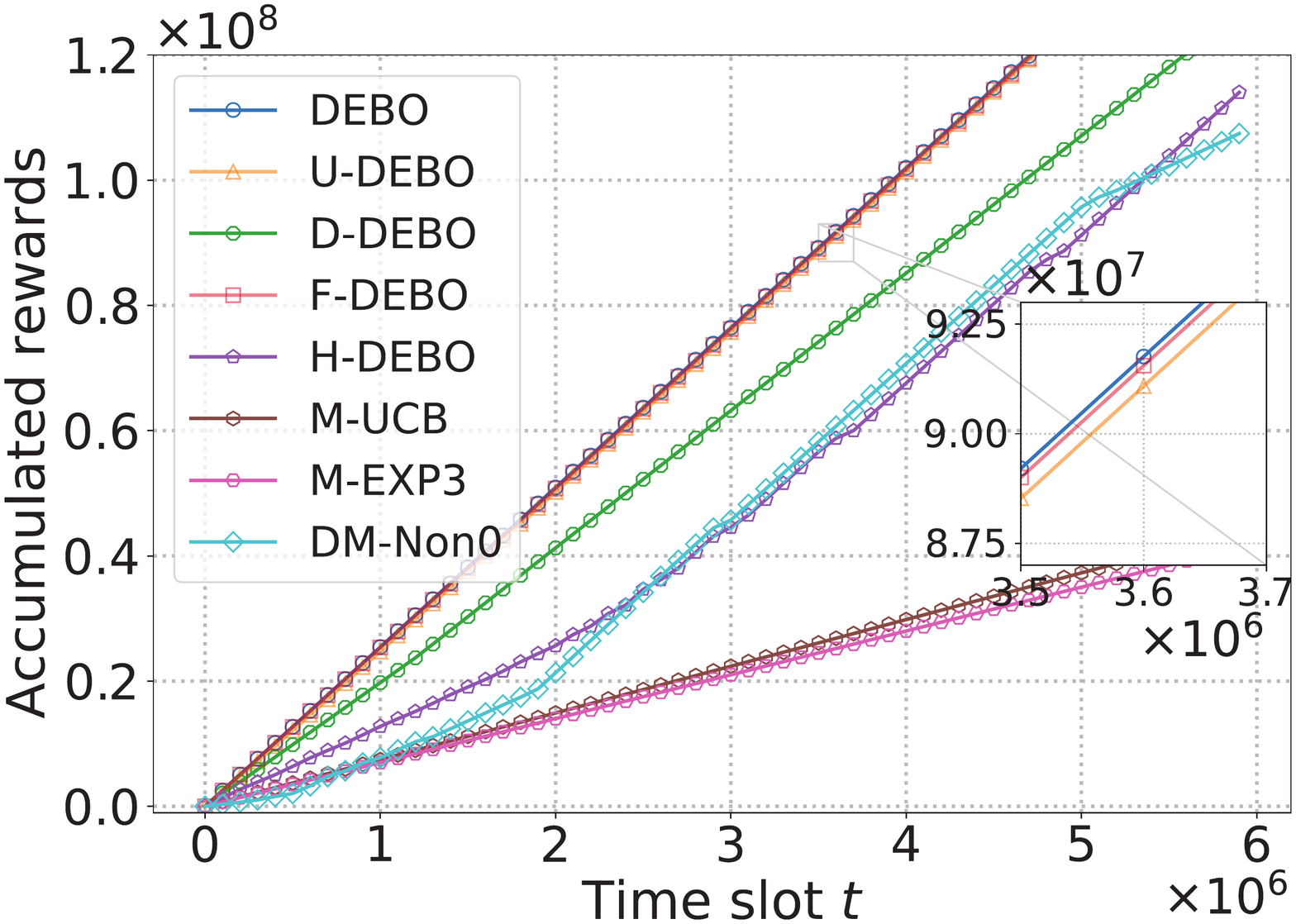}}\ 
   \subfigure[Average rewards]{\includegraphics[width=0.49\linewidth,height=1.3in]{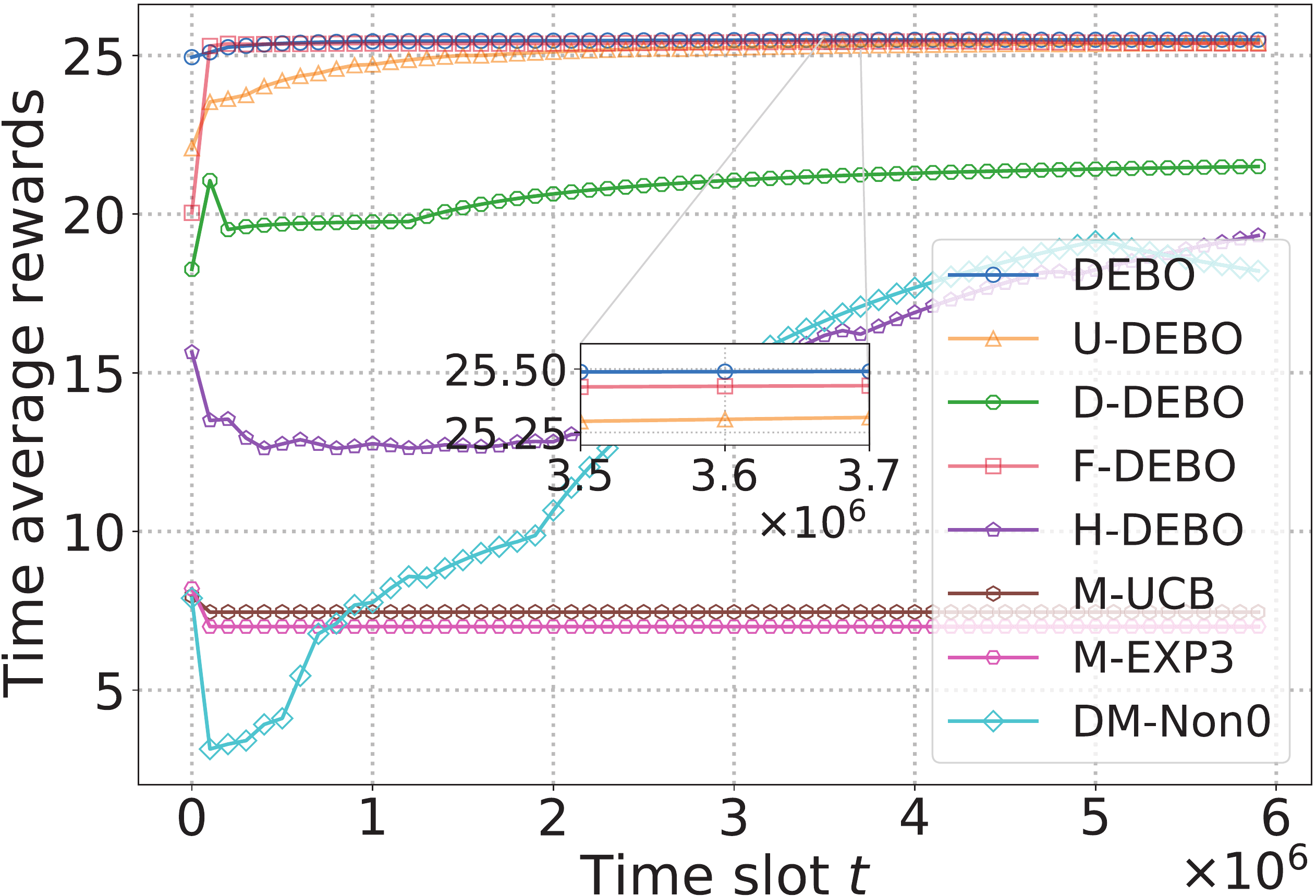}}\\
    \caption{Offloading rewards over time.}
    \label{Fig:CArwd}
  \vspace{-10pt}
\end{figure}

\begin{figure*}[t]
  \centering
   \subfigure[Accumulated regret]{\includegraphics[width=0.27\linewidth,height=1.4in]{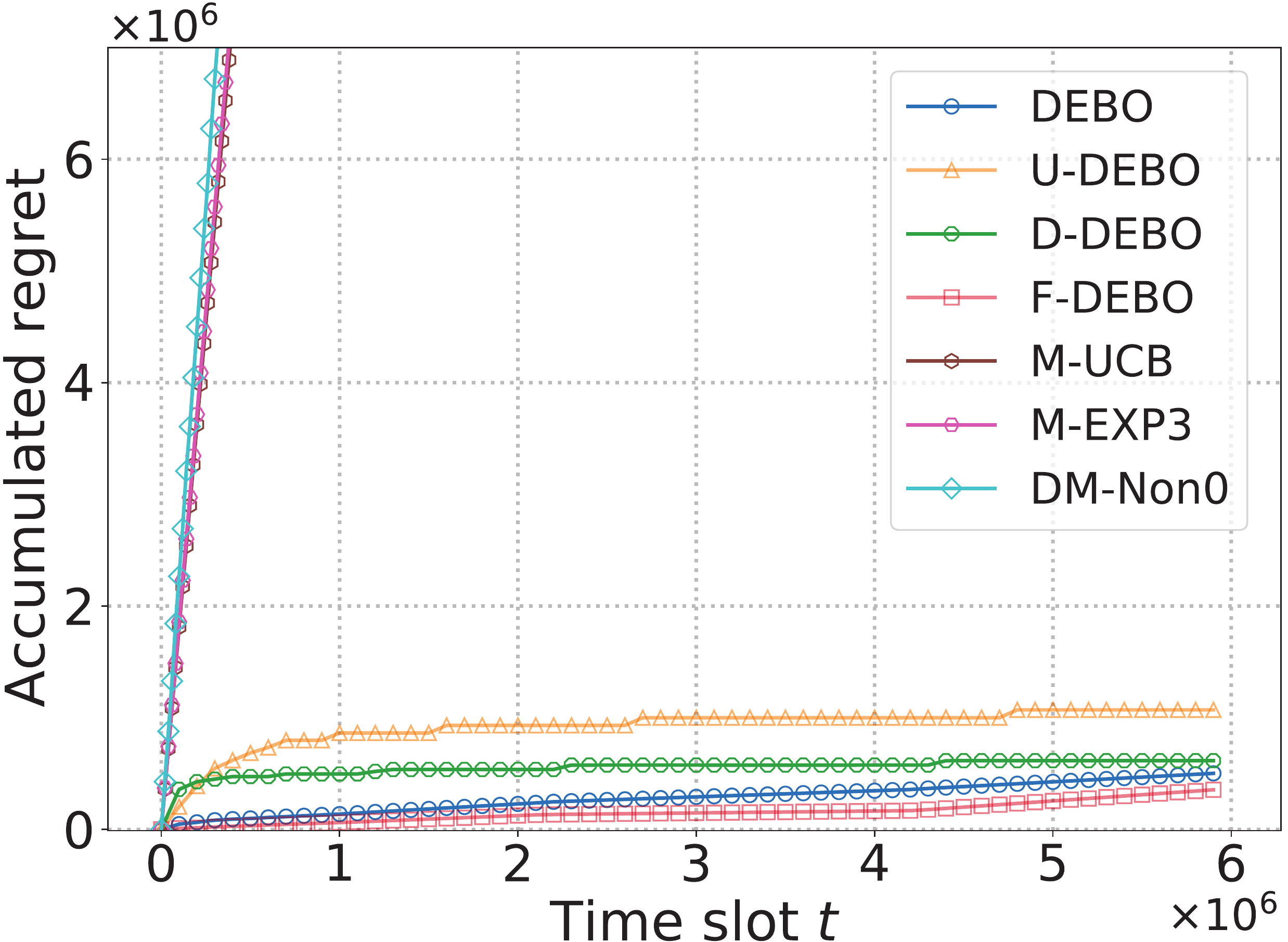}}\ 
   \subfigure[Average regret]{\includegraphics[width=0.27\linewidth,height=1.4in]{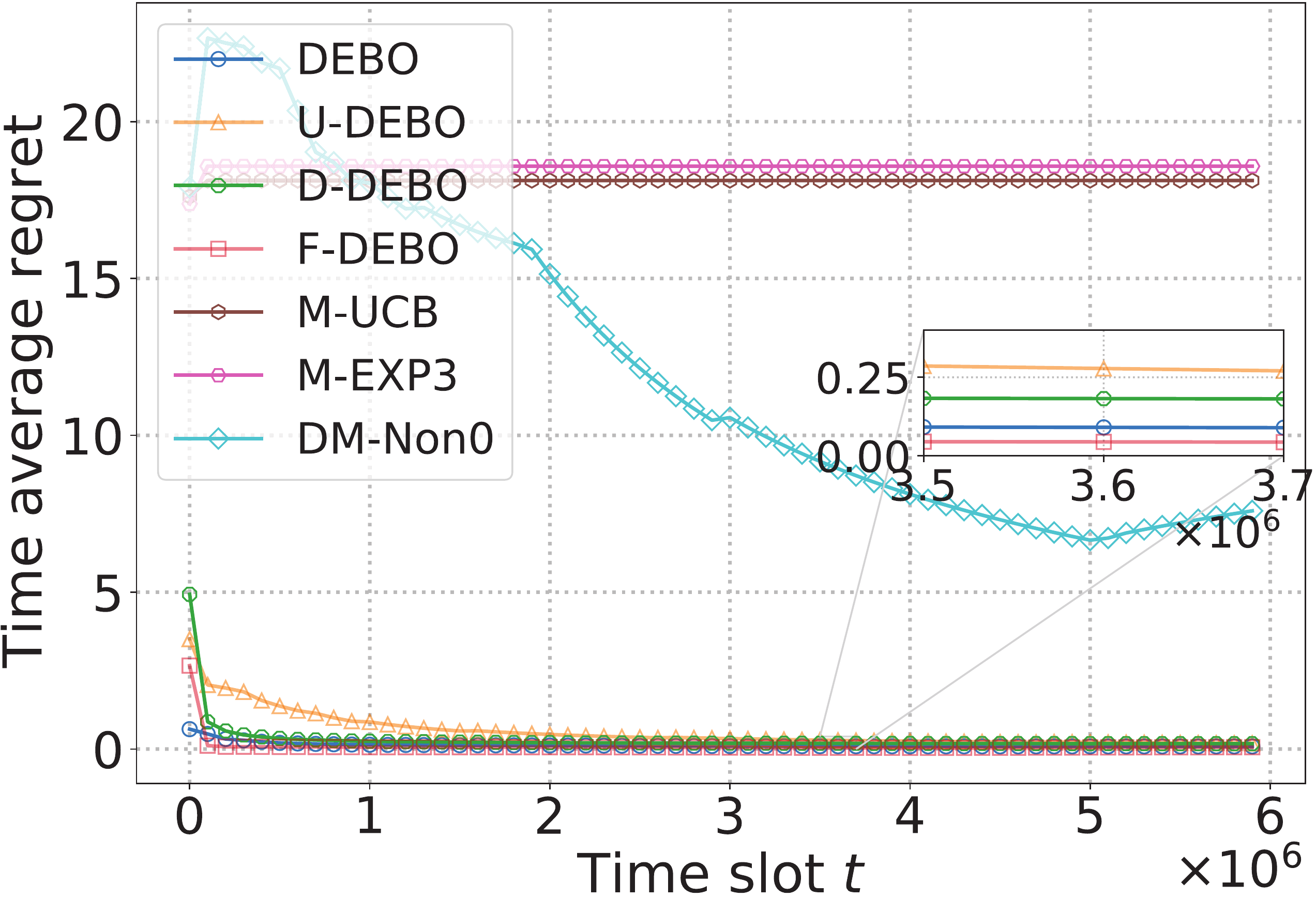}}\
   \subfigure[Reward ratio]{\includegraphics[width=0.27\linewidth,height=1.4in]{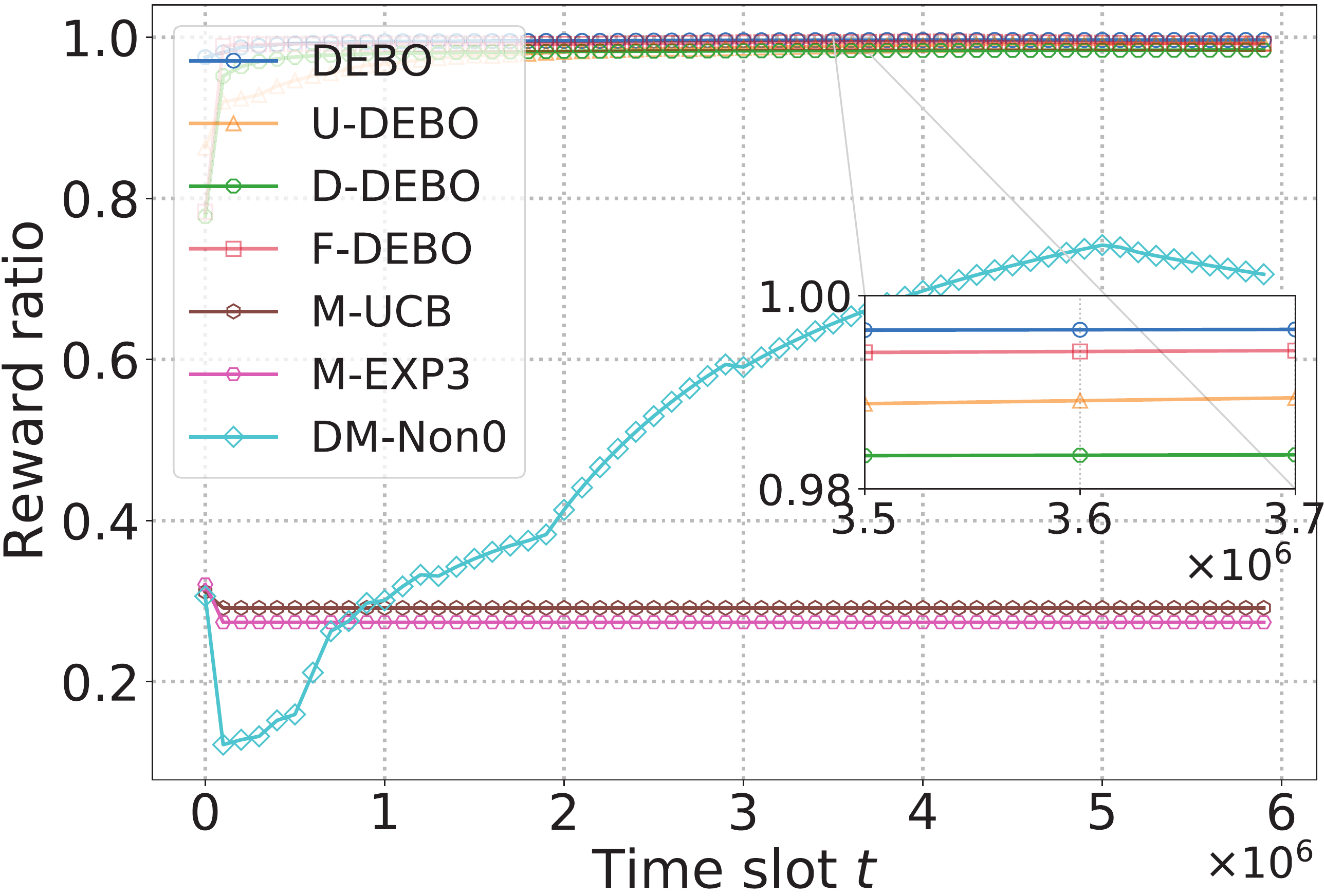}}\\
    \caption{Offloading regret  and  reward ratio over time.}
    \label{Fig:CArt}
\end{figure*}

\begin{figure*}[t]
\centering
\begin{tabular}{cc}
    \begin{minipage}[t]{1.7in}
    \includegraphics[height=1.3in,width=1\columnwidth]{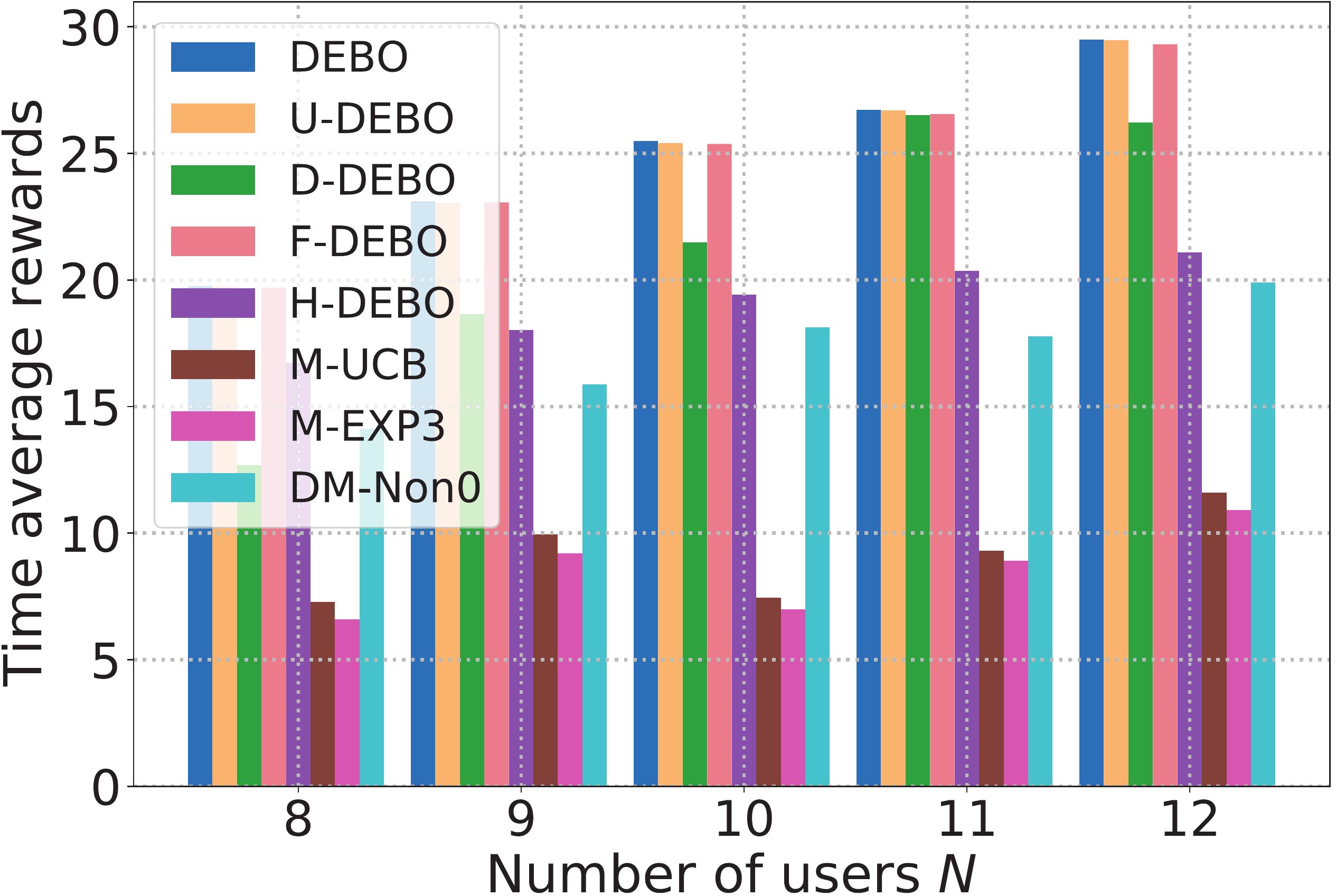}
    \caption{Rewards vs. $N$.}
    \label{Fig:rwdN}
    \end{minipage}

    \begin{minipage}[t]{1.7in}
    \includegraphics[height=1.3in,width=1\columnwidth]{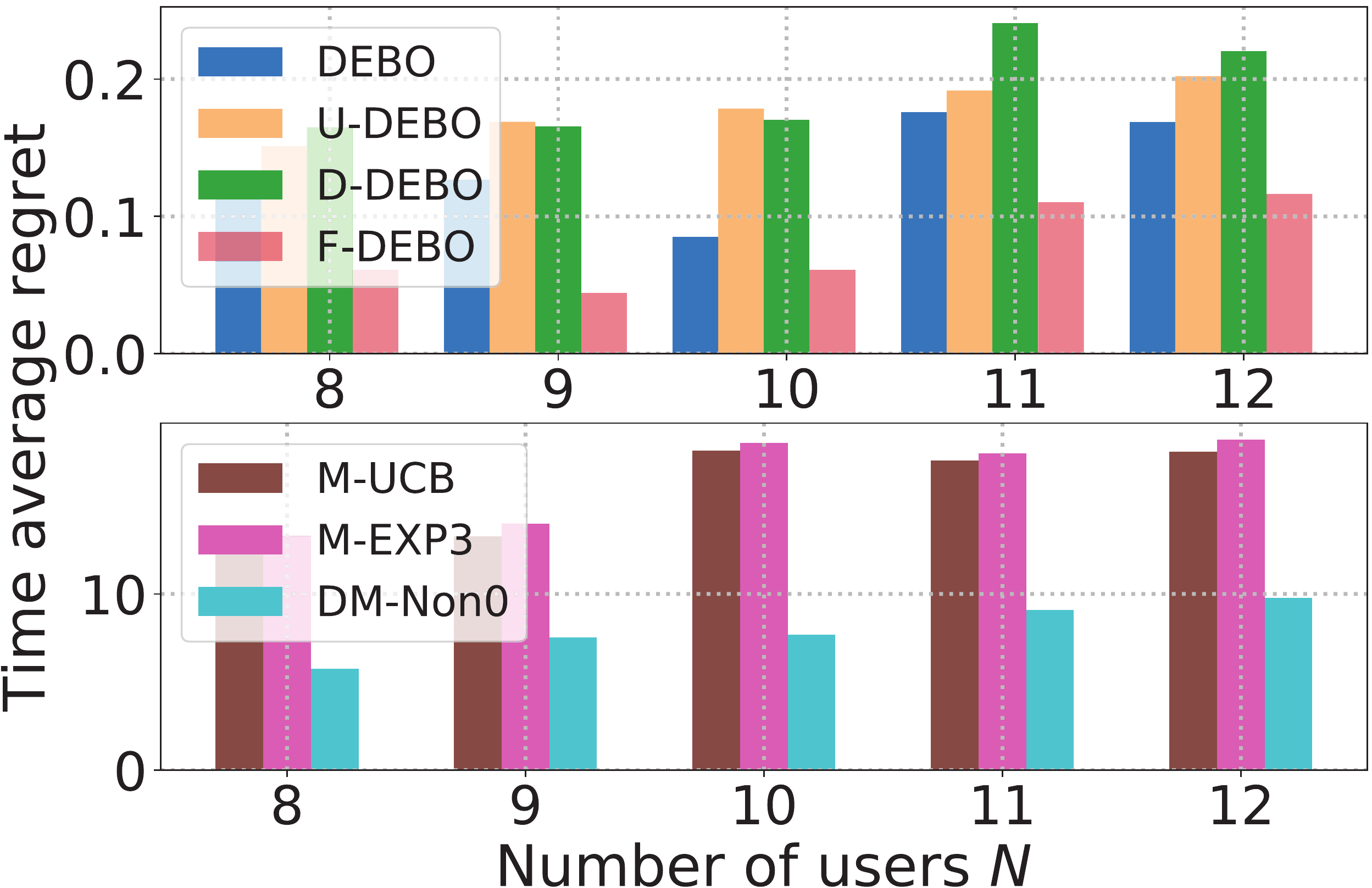}
    \caption{Regret vs. $N$.}
    \label{Fig:rtN}
    \end{minipage}
    
    \begin{minipage}[t]{1.7in}
    \includegraphics[height=1.3in,width=1\columnwidth]{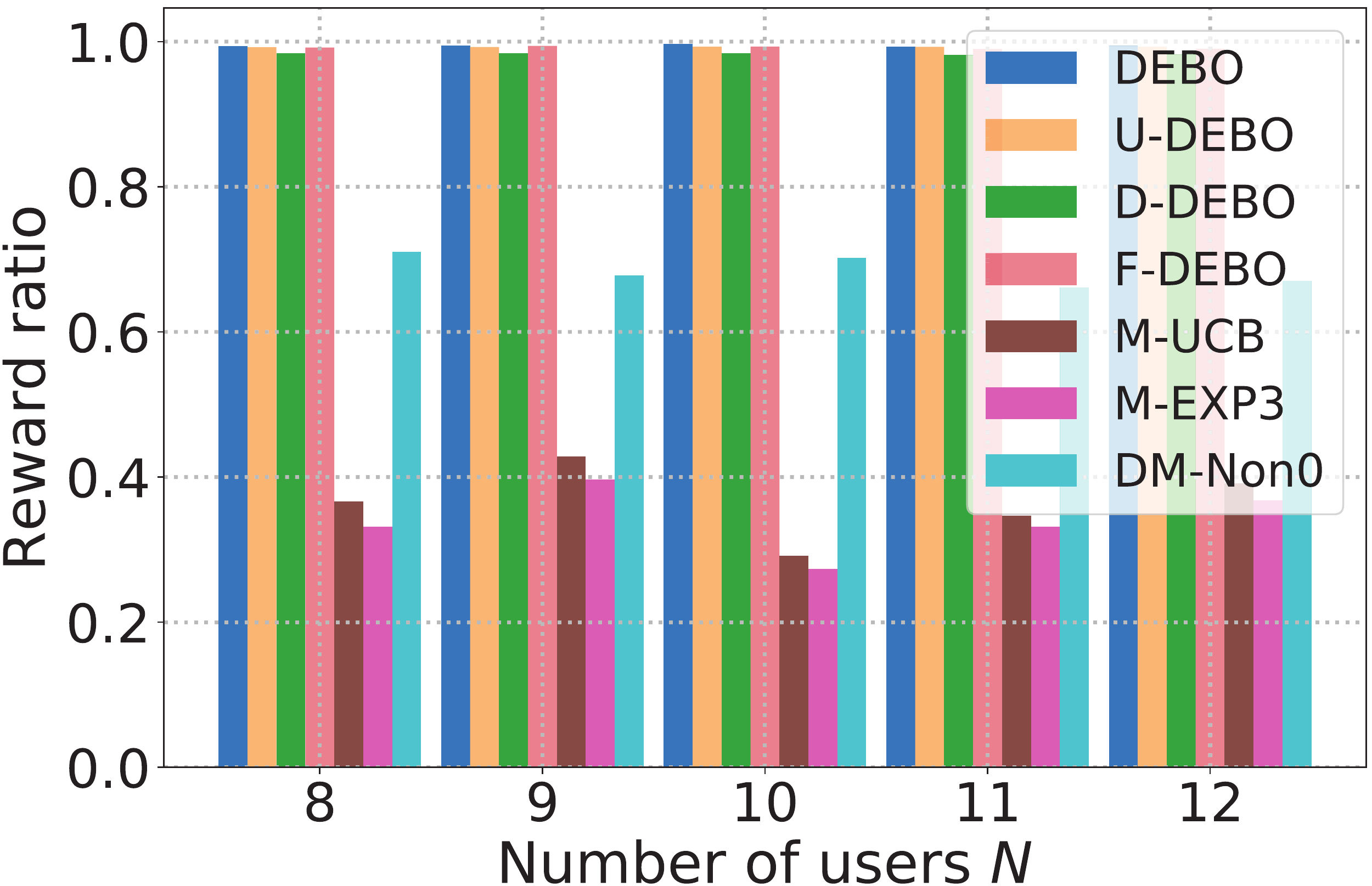}
    \caption{Reward ratio vs. $N$.}
    \label{Fig:ratioN}
    \end{minipage}

    \begin{minipage}[t]{1.7in}
    \includegraphics[height=1.3in,width=1\columnwidth]{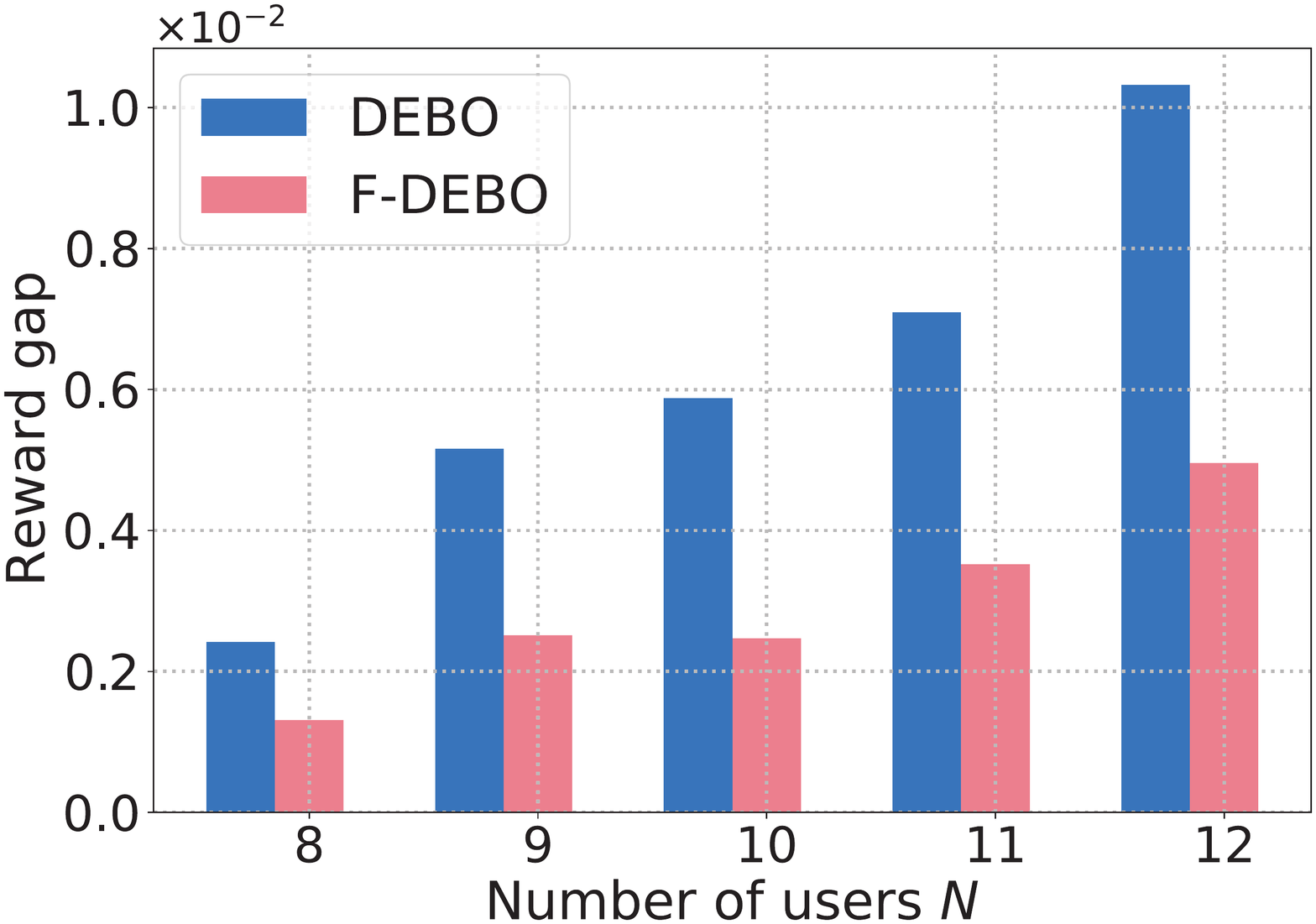}
    \caption{Reward gap vs. $N$.}
    \label{Fig:gapN}
    \end{minipage}
   
\end{tabular}
\end{figure*}

\textbf{Accumulated/average rewards}. The accumulated and time average rewards of our proposed and baseline algorithms are demonstrated in Fig.~\ref{Fig:CArwd}. One can observe that our proposed algorithms outperform the benchmarks as they can always achieve higher accumulated and time average rewards. Particularly, DEBO yields the highest rewards since more information is assumed known in advance, meanwhile U-DEBO and F-DEBO also have satisfactory performances with comparable rewards to DEBO. Besides, D-DEBO leads to slightly lower rewards mainly because dynamic user leaving will cause a reduction in the compound reward $\sum_{i=1}^{N^{(n)}} \mu_{ia_i}$. The average rewards of DEBO and its extensions will stabilize over time, and approach the optimal results, respectively. Moreover, H-DEBO actually attains favorable outcome with large accumulated and average rewards. The baseline method DM-Non0 acquires higher rewards than M-UCB and M-EXP3, i.e., strawman extension of single-user UCB and EXP3 to multi-user scenarios is insufficient to obtain satisfactory performances.

\textbf{Accumulated/average regret and ratio}. The accumulated and time average regrets are shown in Fig.~\ref{Fig:CArt}(a)-(b), where H-DEBO is excluded as we can not derive the optimal assignment to H-OAP. Note that our proposed algorithms give rise to orderly lower regrets than the benchmarks. The regrets of DEBO,  U-DEBO and F-DEBO are negligibly small.  Also, the regret of  D-DEBO is  very  low, which is because the optimal assignment will be changing when we have dynamic user entering or leaving. We then display the ratio between  time average rewards and the optimal result  in  Fig.~\ref{Fig:CArt}(c), which indicates that the ratios of  DEBO and its extensions will converge to steady values close to 1, namely their regrets are sub-linear in terms of the total time horizon $T$. As for M-UCB, M-EXP3 and DM-Non0, there exist performance gaps between their ratios and 1. 

\subsection{Evaluation Results over User Number}
Varying the number of users $N$ from 8 to 12, we obtain  the time average rewards and regret in each case. 

\textbf{Rewards, regret and ratio}. The time average rewards, regret and reward ratio over user number $N$ are shown in Figs.~\ref{Fig:rwdN}-\ref{Fig:ratioN}, respectively. With the increase of $N$, the rewards of all algorithms will tend to increase. Besides, our proposed DEBO, U-DEBO, D-DEBO,  F-DEBO and H-DEBO always achieve better performances than  baseline methods M-UCB, M-EXP3 and DM-Non0 pertaining to  rewards, regret and ratio.

\textbf{Fairness demonstration}. F-DEBO is proposed to ensure a more equitable reward distribution among users by introducing the  proportional fairness. Previously, we have already shown that F-DEBO has comparable rewards to DEBO. Fig.~\ref{Fig:gapN} further displays their maximum gaps of  users' time average  rewards. We can observe that the gap (fairness) is significantly reduced (enhanced) owing to the fairness consideration. In particular,  F-DEBO accomplishes 51.49\% fairness improvement with only sacrificing 0.44\% compound reward on average.

\section{Related Work}\label{Sec:survey}
In this paper, we study the decentralized task offloading in a dynamic MEC system with unknown system-side information. In the following, let us briefly survey the related works.

\textbf{MEC task offloading}. Emerging MEC enables users to offload their computation-intensive tasks to local edge servers~\cite{satyanarayanan2017the}. Chen  \emph{et al.} propose a response updating method to make the offloading decision given the full system-side information~\cite{chen2019task}. Considering system uncertainty, Ouyang \emph{et al.} explore the user-managed service placement by formulating task offloading as a contextual MAB problem, which mainly focuses on  the single-user case~\cite{ouyang2019adaptive}. Following this line, Zhang \emph{et al.} study centralized task offloading under undisclosed task reward and limited server capacity, while they allow a certain degree of capacity violation~\cite{zhang2020an}. To improve the offloading efficiency in the edge computing, a delay-optimal cooperative mobile edge caching scheme is proposed to enhance the MEC service quality with low complexity~\cite{zhang2018cooperative}.  Similarly, the authors in~\cite{zhang2018air} also consider the service caching and  propose a novel service-oriented network slicing approach to efficiently manage the multi-dimensional network resource. So far, decentralized multi-user task offloading with bandit feedbacks still remains open.

\textbf{Decentralized multi-user MAB}. MAB is a representative model for the sequential decision making, where  a decentralized framework is developed  in~\cite{nayyar2018on} for communication based multi-user MAB. Bistritz \emph{et al.} first characterize a fully decentralized bandit policy with heterogeneous rewards based on the theory of unperturbed chain~\cite{bistritz2018distributed}. Later works further apply this framework to the wireless channel allocation~\cite{zafaruddin2019distributed}. However, these existing researches are built on the collision-based reward model, which actually allows an indirect communication signal for decentralized coordination. To the best of our knowledge, only~\cite{magesh2019decentralized} exploits non-zero rewards even collision occurs, but the reward depends on the user number making the same action as long as the number is under a uniform predefined value for all arms.

\textbf{Fairness in MAB}.  Fairness problem in online learning  has attracted a surge of interests. Nevertheless, previous works mainly study arm fairness in single-user scenarios where the selection probability of each arm is supposed to be higher than a predefined proportion~\cite{li2019combinatorial}, or multi-user reward fairness through centralized decision making~\cite{hossain2020fair}. A collision based multi-user reward fairness is also explored in~\cite{bistritz2020my}.

\section{Conclusion}\label{Sec:conclude}
In this paper, we study a fully decentralized task offloading for a dynamic MEC system with the unknown system-side information. We divide the time horizon into epochs and propose DEBO which ensures an $O(\log T)$ regret of the dynamic task offloading in a decentralized manner. On this basis, we also extend DEBO to handle cases such as the unknown reward gap, dynamic user entering or leaving, and  fair reward distribution, where sub-linear regrets are obtained for all extensions. Considering users' heterogeneous resource requests, we further develop H-DEBO which achieves  an $O(\log T)$ regret and satisfactory offloading rewards. Extensive evaluations show the superiority of our proposed algorithms compared to existing benchmarks. 


\begin{appendices}
\appendix
\subsection{Proof of Theorem~\ref{Thm:regret_bound}} \label{App:regret_bound}

Recall that $n_T$ is the last epoch index, then we have:
\begin{equation} \label{Eq:nT}
T \ge \sum_{n=1}^{n_T-1}(T_1+T_2+2^n) \ge 2^{n_T}-2,
\end{equation}
and hence $n_T$ is capped by $\log_2(T+2)$. On the other hand, if the assignment $\bm{a}'$ obtained after the exploration and matching phases is exactly the optimal assignment $\bm{a}^*$, then the exploitation phase will not incur any regret. Because $P_n$ is the error probability that $\bm{a}'$ is not optimal after the exploration and matching phases in epoch $n$, which satisfies  $P_n \le 3NKe^{-n}$ according to Lemma~\ref{Lem:EP}  when setting parameters  $\epsilon$, $T_1$ and $T_2$ as in Theorem~\ref{Thm:regret_bound}. Considering that the regret $\mathcal{R}(T)$ is composed of the regrets in exploration, matching and exploitation phases, respectively, we can attain:
\begin{equation}
\begin{aligned}
\mathcal{R}(T) &\le \sum_{n=1}^{n_T}(T_1N\overline{r}+T_2N\overline{r}+P_n \times 2^n \times N\overline{r})\\
& \le \sum_{n=1}^{n_T}(T_1N\overline{r}+T_2N\overline{r}+3NKe^{-n}\times 2^n \times N\overline{r})\\
& \le (T_1N\overline{r}+T_2N\overline{r})n_T +\frac{6}{e-2}N^2K\overline{r}\\
& \le  (T_1N\overline{r}+T_2N\overline{r})\log_2(T+2)+\frac{6}{e-2}N^2K\overline{r}\\
& \le (T_1N\overline{r}+T_2N\overline{r})\log_2(T+2)+12N^2K\overline{r}\\
&=O(\log_2T).
\end{aligned}
\end{equation}
Proof complete.

\subsection{Proof of Lemma \ref{Lem:EEP}}  \label{App:EEP}
Let $E$ represent the event  that $E = \{\text{there exists}~i,j~\text{such that}~|\tilde{r}_{ij}^{(n)} - \mu_{ij}| >  \Delta\}$. Denote $V_{\min} = \min_{i\in\mathcal{N},j\in\mathcal{K}} V_{ij}$ and and $V_n= nT_1$ after epoch $n$, which are the minimum times a user's tasks are successfully processed by an edge server and the total offloading times in exploration phases, respectively. Also, we use $p_{ij}$ to quantify the probability that user $i$ successfully offloads a task to server $j$. Abiding to RO of Algorithm~\ref{Alg:RO}, we know:
\begin{equation}\label{Eq:pij}
p_{ij} \ge M_j \frac{1}{M}\Bigl(1-\frac{1}{M}\Bigr)^{N-1},
\end{equation}
where the right-hand side is the probability when colliding with other users for a specific resource unit, i.e., different users happen to choose the same integer $H_i$. Provided $V_{\min}$, we cap the error probability of event $E$:
\begin{equation}\label{Eq:EVmin}
\begin{aligned}
\mathrm{Pr}(E|V_{\min}) &\overset{a}{\le} \bigcup_{i =1}^N \bigcup_{j=1}^K \mathrm{Pr}\big(|\tilde{r}_{ij}^{(n)} - \mu_{ij}| >  \Delta\big)\\
& \le NK \max_{i\in \mathcal{N},j \in \mathcal{K}}\mathrm{Pr}\big(|\tilde{r}_{ij}^{(n)} - \mu_{ij}| >  \Delta\big)\\
& \overset{b}{\le} 2NK e^{-\frac{2\Delta^2}{(\overline{r}-\underline{r})^2}V_{\min}},
\end{aligned}
\end{equation}
where $\overset{a}{\le}$ is obtained from the definition of event $E$, and $\overset{b}{\le}$ is based on the Hoeffding's inequality for bounded variables $r_{ij}(t) \in [\underline{r},\overline{r}]$. Moreover, each user has offloaded for exactly $V_n = nT_1$ times after the $n$-th exploration. As $V_{ij}$ records the times when user $i$'s tasks are processed by server $j$ excluding being abandoned, then we have:
\begin{equation}
\begin{aligned}
\mathrm{Pr}\Bigl(V_{\min}<\frac{M_{\min}V_n}{4M}\Bigr) &\overset{c}{=} \mathrm{Pr}\Bigl(\bigcup_{i =1}^N \bigcup_{j=1}^K V_{ij}<\frac{M_{\min}V_n}{4M}\Bigr)\\
& \le \sum_{i=1}^N \sum_{j=1}^K \mathrm{Pr}\Bigl(V_{ij}<\frac{M_{\min}V_n}{4M}\Bigr)\\
& \overset{d}{\le} \sum_{i=1}^N \sum_{j=1}^K e^{-2 \left(p_{ij}-\frac{M_{\min}}{4M}\right)^2V_n},
\end{aligned}
\end{equation}
in which $\overset{c}{=}$ is due to $V_{\min} = \min_{i\in\mathcal{N},j\in\mathcal{K}} V_{ij}$, and $\overset{d}{\le}$ is also raised from the Hoeffding's inequality for Bernoulli random variables. Prior to obtaining the bound of $\mathrm{Pr}\Bigl(V_{\min}<\frac{M_{\min}V_n}{4M}\Bigr)$, we need to discuss the probability $p_{ij}$ in Eq.~\eqref{Eq:pij}. Because $M\ge N$, then $\left(1-\frac{1}{M}\right)^{N-1} \ge \left(1-\frac{1}{N}\right)^{N-1} = \frac{1}{\left(1+\frac{1}{N-1}\right)^{N-1}}$ where $\left(1+\frac{1}{N-1}\right)^{N-1}$ will increase and converge to the natural number $e$ with $N$ increasing, that is $\left(1-\frac{1}{M}\right)^{N-1} \ge \frac{1}{e} > \frac{1}{4}$. Since $M_j \ge M_{\min}$ in Eq.~\eqref{Eq:pij}, then:
\begin{equation} \label{Eq:Vmin}
\begin{aligned}
\mathrm{Pr}\Bigl(V_{\min}<\frac{M_{\min}V_n}{4M}\Bigr) &\le NK e^{-\frac{2M_{\min}^2}{M^2}\left(\left(1-\frac{1}{M}\right)^{N-1}-\frac{1}{4}\right)^2V_n}\\
&\le NK e^{-\frac{2M_{\min}^2}{M^2}\left(\frac{1}{e}-\frac{1}{4}\right)^2V_n}\\
& \overset{f}{\le} NK e^{-\frac{2M_{\min}^2}{81M^2}V_n},
\end{aligned}
\end{equation}
where $\overset{f}{\le}$ uses the fact that $\frac{1}{e}-\frac{1}{4} \ge \frac{1}{9}$. 

At last, we bound the error probability $\mathrm{Pr}(E)$ by combining Eq.~\eqref{Eq:EVmin} and Eq.~\eqref{Eq:Vmin}. In fact, $\mathrm{Pr}(E) = \sum_{V_{\min}} \mathrm{Pr}(E|V_{\min}) \mathrm{Pr}(V_{\min})$, and hence:
\begin{equation} \label{Eq:PrE}
\begin{aligned}
\mathrm{Pr}(E)  & = \sum_{V_{\min}=0}^{\floor*{\frac{M_{\min}V_n}{4M}}} \mathrm{Pr}(E|V_{\min})\mathrm{Pr}(V_{\min})\\
 &+  \sum_{\floor*{\frac{M_{\min}V_n}{4M}}+1}^{V_n}\mathrm{Pr}(E|V_{\min})\mathrm{Pr}(V_{\min})\\
& \le \mathrm{Pr}\Bigl(V_{\min}<\frac{M_{\min}V_n}{4M}\Bigr) + \mathrm{Pr} \Bigl(E|V_{\min} \ge \frac{M_{\min}V_n}{4M}\Bigr)\\
& \le NKe^{-\frac{2M_{\min}^2V_n}{81M^2}} +2NK e^{-\frac{2\Delta^2}{(\overline{r}-\underline{r})^2}\frac{M_{\min}V_n}{4M}}.
\end{aligned}
\end{equation}
Considering that $V_n = nT_1$, and if setting   $T_1 =  \max \Bigl\{\ceil*{\frac{128N^2M(\overline{r}-\underline{r})^2}{9\Delta_{\min}^2 M_{\min}}}, \ceil*{\frac{81M^2}{2M_{\min}^2}}\Bigr\}$ while letting $\Delta = \frac{3\Delta_{\min}}{8N}$, we immediately attain the exploration error probability in Eq.~\eqref{Eq:EEP}.

\subsection{Proofs of Lemmas~\ref{Lem:eCS}-\ref{Lem:EP}}
\subsubsection{Proof of  Lemma~\ref{Lem:eCS}}  \label{App:eCS}
$\epsilon$-CS is proved by induction. Initially, all users are unassigned as $a_i = 0, \forall i \in \mathcal{N}$, $\epsilon$-CS certainly holds. Suppose that the current assignment satisfies $\epsilon$-CS, we demonstrate that this is still true if a new user-resource unit match is added.

First, assume that user $i$ sends a bid (offloads a task) to server $j(m^*)$ for requesting resource unit $m^*$, and $m^*$ is not assigned to any other users before. Denote $\bm{B}$ and $\bm{B}'$ as the bid matrices before and after the assignment. According to DAuction in  Algorithm~\ref{Alg:DAuction}, we have $B'_{im^*} = R_{im^*} -(R_{im'}-B_{im'}) + \epsilon$ (Line 10), which is also the highest bid among all received bids for the resource unit $m^*$. Then, we attain:
\begin{equation}
\begin{aligned}
R_{im^*} -\eta'_{m^*} &=  R_{im^*}-B'_{im^*}\\
 &= (R_{im'}-B_{im'}) - \epsilon\\
& = \max_{m\notin M(m^*)}\{R_{im}-B_{im}\} -\epsilon\\
&\overset{a}{\ge} \max_{m\notin M(m^*)}\{R_{im}-\eta'_{m}\} - \epsilon\\
& \overset{b}{=}\max_{m\notin M(m^*)}\{R_{im}-\tilde{\eta}'_{j(m)}\} - \epsilon,
\end{aligned}
\end{equation}
where $ \overset{a}{\ge}$ is because each $\eta_{m}$ is non-decreasing when a new match is added, and $\overset{b}{=}$ is because $\tilde{\eta}'_{j(m)} = \min_{m\in M(m)}\{\eta'_m\}$. Besides, since $\tilde{\eta}'_{j(m^*)} \le \eta'_{m^*}$, then $R_{im^*} - \tilde{\eta}'_{j(m^*)}  \ge R_{im^*} -\eta'_{m^*}$. Also, $\tilde{\eta}'_{j(m^*)}=\tilde{\eta}'_{j(m)}, \forall m\in M(m^*)$, which leads to $R_{im^*} - \tilde{\eta}'_{j(m^*)} = R_{im} - \tilde{\eta}'_{j(m)}, \forall m \in M(m)$ owing to $R_{im^*} = R_{im}, \forall m \in M(m^*)$. As a result, $R_{im} - \tilde{\eta}_{j(m)} \ge \max_{m} \{R_{im}-\tilde{\eta}_{j(m)}\} - \epsilon$ or $\epsilon$-CS stands.

Second, we discuss another case where the user $i$-resource unit $m^*$ match already exists, and this match remains unchanged after a new round of bidding from other users. In the following, we use notations, like $\bm{B}'', \eta''$, to denote values of corresponding parameters when the match,  user $i$-resource unit $m^*$,  first joins in the assignment. Therefore,
\begin{equation}
\begin{aligned}
R_{im^*} -\tilde{\eta}''_{j(m^*)} &\ge  R_{im^*}-\eta''_{m^*}\\
&=  R_{im^*}-B''_{im^*}\\
&=(R_{im'}-B''_{im'}) - \epsilon \\
&= \max_{m\notin M(m^*)}\{R_{im}-B''_{im}\} -\epsilon\\
& \overset{c}{=} \max_{m\notin M(m^*)}\{R_{im}-\eta''_{m}\} -\epsilon \\
&\overset{d}{\ge} \max_{m\notin M(m^*)}\{R_{im}-\eta_{m}\} - \epsilon,
\end{aligned}
\end{equation}
where $\overset{c}{=}$ is from Line 10 in Algorithm~\ref{Alg:DAuction}, and $\overset{d}{\ge}$ is because $\eta_{m}$ is non-decreasing. On the other side,  $\tilde{\eta}''_{j(m^*)}$ does not change as user $i$ always wins the auction for unit $m^*$, then $R_{im^*} -\tilde{\eta}''_{j(m^*)} =R_{im^*} -\tilde{\eta}_{j(m^*)}$, i.e., $\epsilon$-CS holds for $m^*$. For any other $m\in M(m^*)$, since  $\tilde{\eta}_{j(m)}=\tilde{\eta}_{j(m^*)}$, we have $R_{im^*} - \tilde{\eta}_{j(m^*)} = R_{im} - \tilde{\eta}_{j(m)}$ and $\epsilon$-CS still satisfies. 

Overall, the final assignment $\bm{a}$ meets $\epsilon$-CS.

\subsubsection{Proof of Lemma~\ref{Lem:ME}} \label{App:ME}
Similar to the proof of Lemma~\ref{Lem:eCS}, we still use resource unit assignment $\bm{a}$ and rewards $\bm{R}$ for better illustration, which are equivalent to the server assignment $\bm{a}'$ in Algorithm~\ref{Alg:DAuction} and the estimated rewards $\tilde{\bm{r}}^{(n)}$.

To stating with, we show if $K\epsilon < \tilde{\Delta}_{\min}$, then the assignment $\bm{a}$ (namely $\bm{a}'$) is optimal under $\bm{R}$ (namely $\tilde{\bm{r}}^{(n)}$). Suppose in the contrast, assignment $\bm{a}$ is not optimal, then we can choose a subset from $\bm{a}$ different from the subset from the optimal assignment $\bm{a}^{(n)}$, that involves a one-to-one match from a $K$-user subset $\mathcal{N}'$ to the edge server set $\mathcal{K}$. From the $\epsilon$-CS in Lemma~\ref{Lem:eCS}, we know that:
\begin{displaymath}
\sum_{i \in \mathcal{N}'} (R_{ia_i} - \tilde{\eta}_{j(a_i)}) \ge \sum_{i \in \mathcal{N}'} \left(\max_{m} \{R_{im}-\tilde{\eta}_{j(m)}\} - \epsilon\right),
\end{displaymath}
that is, $\sum_{i \in \mathcal{N}'} R_{ia_i} \ge  \sum_{i \in \mathcal{N}'} (\max_{m} \{R_{im}\}) -K\epsilon \ge   \sum_{i \in \mathcal{N}'} R_{ia^{(n)}_i} -K\epsilon$ due to the one-to-one match. Since $K\epsilon <\tilde{\Delta}_{\min}$, which means $a_i, i\in \mathcal{N}'$ is the same as the optimal assignment $a^{(n)}_i, i\in \mathcal{N}'$, which is a contradiction. For the case if we can only select a subset of edge servers, it follows a similar vein because when $\bm{a}$ is suboptimal, one could always pick a $K$-user subset to demonstrate a contrast. Otherwise, the reward gap between $\bm{a}$ and $\bm{a}^{(n)}$ will be lower than $K\epsilon$ when no such $K$-user subset exists, i.e.,  $\bm{a}$ is still optimal.

As for Eq.~\eqref{Eq:ME}, we can leverage $\epsilon$-CS by replacing $\tilde{\eta}_{j(m)}$ with $\eta_{m}$, that is $R_{im} - \eta_{m} \ge \max_{m} \{R_{im}-\eta_{m}\} - \epsilon$, while the proof is similar to that of Lemma~\ref{Lem:eCS}.  For those resource units $m$ with no bids, we have $\eta_{m}=0$. Since
\begin{displaymath}
\sum_{i \in \mathcal{N}} (R_{ia_i} - \eta_{a_i}) \ge \sum_{i \in \mathcal{N}} \left(\max_{m} \{R_{im}-\eta_{m}\} - \epsilon\right),
\end{displaymath}
then $\sum_{i \in \mathcal{N}} R_{ia_i} \ge  \sum_{m=1}^M \eta_m + \sum_{i \in \mathcal{N}} \left(\max_{m} \{R_{im}-\eta_{m}\} - \epsilon\right) =  \sum_{i \in \mathcal{N}} (\max_{m} \{R_{im}\}) -N\epsilon$.

As for the termination time, it follows Theorem~4 in~\cite{naparstek2014fully}. For any unassigned user $i$, each time the bid for a resource unit will increase by at least $\epsilon$, or the net utility on the best resource unit will decrease by at least $\epsilon$. After $T'_i = M+\frac{1}{\epsilon}\sum_{m=1}^M R_{im}$, we have $R_{im} - B_{im} < 0, \forall m$, implying $B_{im}>0, \forall m$. Therefore, user $i$ will remain assigned. Sum up all $T_i'$ and use the fact that $R_{im} < \overline{r}$, then we complete the proof.

\subsubsection{Proof of Lemma~\ref{Lem:EP}} \label{App:EP}
First of all, let us illustrate the generation of error before the exploitation phase. In particular, this error stems from  the reward estimation error and the matching error. Based on Lemma~\ref{Lem:EEP}, we know that $|\tilde{r}_{ij}^{(n)} - \mu_{ij}| \le  \frac{3\Delta_{\min}}{8N}, \forall i \in \mathcal{N}, j \in \mathcal{K}$ with probability at least $1-3NKe^{-n}$ in epoch $n$. Denote $\delta_{ij} = \tilde{r}_{ij}^{(n)}-\mu_{ij}$, and $\bm{a}^*$ as the optimal assignment defined before, then:
\begin{equation}
\sum_{i=1}^N \tilde{r}^{(n)}_{ia^*_i} =\sum_{i=1}^N (\mu_{ia^*_i}+\delta_{ia^*_i}) \ge \sum_{i=1}^N \mu_{ia^*_i} -  \frac{3\Delta_{\min}}{8}.
\end{equation}
Furthermore, for the obtained assignment $\bm{a}'$ in DAuction:
\begin{equation}
\sum_{i=1}^N \tilde{r}^{(n)}_{ia'_i} =\sum_{i=1}^N (\mu_{ia'_i}+\delta_{ia'_i}) \le \sum_{i=1}^N \mu_{ia'_i} + \frac{3\Delta_{\min}}{8}.
\end{equation}
On the other hand, Lemmas~\ref{Lem:ME} states that when  $\epsilon = \max \{\frac{\Delta_{\min}}{5N}, \frac{\delta_{\min}}{K}-\frac{3\Delta_{\min}}{4NK}\}$, the gap between 
$\sum_{i=1}^N \tilde{r}^{(n)}_{ia'_i}$ and $\sum_{i=1}^N \tilde{r}^{(n)}_{ia^{(n)}_i}$ is at most $\frac{\Delta_{\min}}{5}$ within $T_2$ time slots.
Hence:
\begin{equation}
\begin{aligned}
\sum_{i=1}^N \mu_{ia^*_i} - \sum_{i=1}^N \mu_{ia'_i} & \le \sum_{i=1}^N \tilde{r}^{(n)}_{ia^*_i} +  \frac{3\Delta_{\min}}{8} - \Bigl(\sum_{i=1}^N \tilde{r}^{(n)}_{ia'_i}  -  \frac{3\Delta_{\min}}{8}\Bigr)\\
& = \sum_{i=1}^N \tilde{r}^{(n)}_{ia^*_i} - \sum_{i=1}^N \tilde{r}^{(n)}_{ia'_i} +\frac{3\Delta_{\min}}{4}\\
& \overset{a}{\le}  \sum_{i=1}^N \tilde{r}^{(n)}_{ia^{(n)}_i} -  \sum_{i=1}^N \tilde{r}^{(n)}_{ia'_i} +\frac{3\Delta_{\min}}{4}\\
& \le \frac{\Delta_{\min}}{5} + \frac{3\Delta_{\min}}{4}\\
& < \Delta_{\min},
\end{aligned}
\end{equation}
where $\overset{a}{\le}$ is because $\bm{a}^{(n)}$ is optimal under $\tilde{\bm{r}}^{(n)}$. Recall from the definition of $ \Delta_{\min}$ which is the compound reward gap between the optimal and the highest suboptimal assignments. In other words,  $\bm{a}' = \bm{a}^*$ with probability at least $1-3NKe^{-n}$, i.e., $P_n \le 3NKe^{-n}$.

\subsection{Proofs of Theorems~\ref{Thm:unknownR} and~\ref{Thm:DMEC}}
\subsubsection{Proof of Theorem~\ref{Thm:unknownR}} \label{App:unknownR}
This proof is similar to that of Theorem~\ref{Thm:regret_bound}, while we mainly unravel  the bound of error probability $P_n$ that assignment $\bm{a}'$ it not optimal. Similar to Eq.~\eqref{Eq:nT}, we have $n_T \le \log_2 (T+2)$. In fact, Eq.~\eqref{Eq:unknownR} amounts to three parts, that is the exploration regret $\mathcal{R}_O(T)$, matching regret $\mathcal{R}_M (T)$ and exploitation regret $\mathcal{R}_I(T)$, respectively, which will be derived sequentially.

For  $\mathcal{R}_O(T)$, we have:
\begin{displaymath}
\begin{aligned}
\mathcal{R}_O(T) \le \sum_{n=1}^{n_T} N\overline{r} T_1^{(n)} \le N\overline{r} T_1^{(n_T)}  n_T.
\end{aligned}
\end{displaymath}
Substituting $T_1^{(n_T)}$ and $n_T \le \log_2 (T+2)$, we obtain $\mathcal{R}_O(T) \le \ceil*{c_1 \log_2^{\vartheta}(T+2)}N\overline{r} \log_2(T+2) $. Analogously, we can attain the matching regret $\mathcal{R}_M(T) \le \ceil*{MN+\log_2^{\vartheta}(T+2) MN\overline{r}/c_0} N\overline{r} \log_2(T+2)$.

We now focus on the exploitation regret $\mathcal{R}_I(T)$. Following the same approach of deriving Eq.~\eqref{Eq:PrE} in Lemma~\ref{Lem:EEP}, the exploration error probability $\mathrm{Pr}(E)$ in Eq.~\eqref{Eq:EEP} satisfies: 
\begin{equation}
\mathrm{Pr}(E)  \le NKe^{-\frac{2M_{\min}^2}{81M^2}V_n} +2NK e^{-\frac{9\Delta_{\min}^2 M_{\min}}{128N^2M (\overline{r}-\underline{r})^2}V_n},
\end{equation}
where $V_n =\sum_{l=1}^n T_{1}^{(l)}$. Denote $c_2 = \min \Bigl \{ \frac{2M_{\min}^2}{81M^2}, \frac{9\Delta_{\min}^2 M_{\min}}{128N^2M (\overline{r}-\underline{r})^2}\Bigr\}$, and then $\mathrm{Pr}(E) \le 3NKe^{-c_2\sum_{l=1}^n T_{1}^{(l)}}$. As $T_{1}^{(n)} = \ceil*{c_1 n^{\vartheta}} \ge c_1 n^{\vartheta}$, we consider $c_1 c_2  \sum_{l=1}^{n_1}  l^{\vartheta} \ge n_1$ in and after epoch $n_1$. Because $\sum_{l=1}^{n_1}  l^{\vartheta} \ge \int_{1}^{n_1+1}(x-1)^{\vartheta}dx =\frac{1}{\vartheta+1}n_1^{\vartheta+1}\ge \frac{1}{2}n_1^{\vartheta+1}$, then if setting $n_1 =\ceil*{[2/(c_1c_2)]^{1/\vartheta}}$, we attain $c_2 \sum_{l=1}^n T_{1}^{(l)}\ge n, \forall n \ge n_1$. As a result, $\mathrm{Pr}(E) \le 3NKe^{-n}, \forall n \ge n_1$. Similarly, since $\epsilon^{(n)} = c_0 n^{-\vartheta} \rightarrow 0$, there must exist a finite integer $n_2$ such that $\epsilon^{(n)} \le  \max \{\frac{\Delta_{\min}}{5N}, \frac{\delta_{\min}}{K}-\frac{3\Delta_{\min}}{4NK}\}, \forall n \ge n_2$. Set $n_0 = \max \{n_1,n_2\}$.  In line with Lemma~\ref{Lem:EP}, the error probability $P_n \le 3NKe^{-n}, \forall n \ge n_0$. Therefore:
\begin{displaymath}
\begin{aligned}
\mathcal{R}_I(T) & \le \sum_{n=1}^{n_T} N \overline{r}\times 2^n \times P_n  \\
& \le  \sum_{n=1}^{n_0}  N \overline{r}\times 2^n +  \sum_{n=n_0}^{+\infty}  3N^2K \overline{r}\times \left(\frac{2}{e}\right)^n\\
& = N\overline{r} (2^{n_0}-1) + \frac{3N^2Ke}{e-2} \left(\frac{2}{e}\right)^{n_0}\\
& \le N\overline{r} (2^{n_0}-1) + 12 N^2 K.
\end{aligned}
\end{displaymath}
By summing up $\mathcal{R}_O(T)$, $\mathcal{R}_M(T)$ and $\mathcal{R}_I(T)$, we complete the proof.

\subsubsection{Proof of Theorem~\ref{Thm:DMEC}}  \label{App:DMEC}
The leaving case still has $O(\log_2 T)$ as analyzed above, so we focus on the entering case. From Lemma~\ref{Lem:EP}, we have $P_n \le 3N^{(n)}Ke^{-(n+1-n')}, \forall n \ge n'$.  Since $n' \le O(\log_2 T^{\zeta})$, there exists a $n_0 \le O(\log_2 T^{\zeta}), n_0 \ge n'$ such that $3N^{(n)}Ke^{-(n+1-n')} \times 2^n \sim O(1), \forall n \ge n_0$. As $n_T \le O(\log_2 T)$ proved in Theorem~\ref{Thm:regret_bound},  the exploration regret and matching regret are $\mathcal{R}_O(T) \le O(\log_2 T)$ and $\mathcal{R}_M(T) \le O(\log_2 T)$, respectively. For the exploitation regret $\mathcal{R}_I(T)$, we attain:
\begin{equation}
\begin{aligned}
\mathcal{R}_I(T) & \le \sum_{n=1}^{n_T} N\overline{r} \times 2^n \times P_n\\
& \le  \sum_{n=1}^{n_0}  N \overline{r}\times 2^n +  \sum_{n=n_0}^{n_T} O(1) \\
&\le    N \overline{r}\times 2^{O(\log_2 T^{\zeta})} + O(\log_2 T).\\
& = O(T^{\zeta}).
\end{aligned}
\end{equation}
Therefore, $\mathcal{R}(T) \le O(T^{\zeta})$.

\subsection{Proof of Lemma~\ref{Lem:logE}} \label{App:logE}
Since  $|\tilde{r}_{ij}^{(n)} - \mu_{ij}| \le \Delta$, we have: $ \ln \left(1+ \beta \tilde{r}_{ij}^{(n)} \right)- \ln (1+ \beta \mu_{ij}) \le \ln [1+ \beta (\mu_{ij}+\Delta)] - \ln (1+ \beta \mu_{ij}) = \ln \frac{1+ \beta (\mu_{ij}+\Delta)}{1+ \beta \mu_{ij}} \le  \ln \left( 1+ \frac{\beta \Delta}{1+ \beta \underline{r}}\right)  \overset{a}{\le} \frac{\beta \Delta}{1+ \beta \underline{r}}$, where $ \overset{a}{\le}$ is due to $\ln (1+x) \le x, \forall x \ge 0$.

On the other hand, $\ln \left(1+ \beta \tilde{r}_{ij}^{(n)} \right)- \ln (1+ \beta \mu_{ij})  \ge \ln [1+ \beta (\mu_{ij}- \Delta)] - \ln (1+ \beta \mu_{ij})  = \ln \frac{1+ \beta (\mu_{ij}-\Delta)}{1+ \beta \mu_{ij}}\overset{b}{\ge} -\frac{\beta \Delta}{1+ \beta \mu_{ij} - \beta \Delta} \ge -\frac{\beta \Delta}{1+ \beta \underline{r} - \beta \Delta} \overset{c}{\ge} -   \frac {4\beta \Delta}{3(1+\beta \underline{r})}$. Here, $\overset{c}{\ge}$ is because $\Delta < \frac{1+\beta \underline{r}}{4 \beta}$ which implies $1+ \beta \underline{r} - \beta \Delta \ge \frac{3}{4}(1+\beta \underline{r})$. As for $\overset{b}{\ge}$, it holds since $\ln x \ge \frac{x-1}{x}, \forall x >0$, and we let $x =\frac{1+ \beta (\mu_{ij}-\Delta)}{1+ \beta \mu_{ij}}$. For completeness, we briefly show why $\ln x \ge \frac{x-1}{x}, \forall x >0$. Denote $h(x) = x \ln x -x+1$. Therefore, $h'(x) = \ln x$, i.e., $h(x)$ decreases in $(0,1]$ and increases in $(1, +\infty)$. Since $h(1)= 0$, we have $x \ln x -x+1 \ge 0$, and hence $\ln x \ge \frac{x-1}{x}$ if re-arranging each term. In total, we can derive Eq.~\eqref{Eq:logE} by combining these two cases.

\subsection{Proof of Theorem~\ref{Thm:heter_regret}} \label{App:heter_regret}
In the first place, we bound the exploration error probability. Denote $E =\{$there exists $i$, $j$ such that $|\tilde{r}_{ij}^{(n)}-\mu_{ij}| > \frac{\delta'_{\min}}{5\overline{M}_{\max}}\}$. After the $n$-th exploration according to the group offloading, any user $i$ has perceived reward observations from server $j$ for at least $V_{\min} = \frac{T_1}{K}n =\ceil*{\frac{25 \overline{M}_{\max}^2(\overline{r}-\underline{r})^2}{2 (\delta'_{\min})^2}}n$ times when $N_g \le K$, which is also true when $N_g > K$. Therefore, the exploration error probability satisfies:
\begin{equation}\label{Eq:HEVmin}
\begin{aligned}
\mathrm{Pr}(E|V_{\min}) & \le \bigcup_{i =1}^N \bigcup_{j=1}^K \mathrm{Pr}\Bigl(|\tilde{r}_{ij}^{(n)} - \mu_{ij}| >  \frac{\delta'_{\min}}{5\overline{M}_{\max}} \Bigl)\\
& \le NK \max_{i\in \mathcal{N},j \in \mathcal{K}}\mathrm{Pr}\Bigl(|\tilde{r}_{ij}^{(n)} - \mu_{ij}| >  \frac{\delta'_{\min}}{5\overline{M}_{\max}} \Bigr)\\
& \overset{a}{\le} 2NK e^{-\frac{2 (\delta'_{\min})^2}{25\overline{M}_{\max}^2 (\overline{r}-\underline{r})^2}V_{\min}}\\
&\le 2NK e^{-n},
\end{aligned}
\end{equation}
where $\overset{a}{\le}$ is based on the Hoeffding's inequality.

For the matching phase, we have $|\tilde{r}_{ij}^{(n)} - \mu_{ij}| \le  \frac{\delta'_{\min}}{5\overline{M}_{\max}}$ with probability at least $1-2NK e^{-n}$. Denote $\delta_{ij}=\tilde{r}_{ij}^{(n)} - \mu_{ij}$, and $\bm{I}^{(\mu)}$ as the indicator which induces the assignment $\bm{a}^{(\mu)}$ to H-OAP of Eq.~\eqref{Eq:opt_heter} given expected  rewards $\{\mu_{ij}\}$. Regarding the subproblem Eq.~\eqref{Eq:knapsack} corresponding to the edge server $j$ under the optimal assignment,  let $OPT_j(\bm{a}^{(\mu)})= \sum_{i=1}^N \Delta \mu_{ij} =\sum_{i=1}^N (\mu_{ij}-\mu_{iI^{(\mu)}_i})$ and $OPT'_j(\bm{a}^{(\mu)}) = \sum_{i=1}^N \Delta \tilde{r}_{ij}^{(n)} = \sum_{i=1}^N (\tilde{r}_{ij}^{(n)}-\tilde{r}_{iI^{(\mu)}_i}^{(n)})$.  Hence, $OPT_j(\bm{a}^{(\mu)})-OPT'_j(\bm{a}^{(\mu)}) \overset{b}{\le} \overline{M}_{\max} \max \{|\delta_{ij}|+ |\delta_{iI^{(\mu)}_i}|\} \le \frac{2\delta'_{\min}}{5}$  where $\overset{b}{\le}$ is because any server can accommodate at most $\overline{M}_{\max}$ users. Under the obtained assignment $\bm{a}'$, we similarly denote  $OPT_j(\bm{a}')= \sum_{i=1}^N \Delta \mu_{ij} =\sum_{i=1}^N (\mu_{ij}-\mu_{iI_i})$ and $OPT'_j(\bm{a}') = \sum_{i=1}^N \Delta \tilde{r}_{ij}^{(n)} = \sum_{i=1}^N (\tilde{r}_{ij}^{(n)}-\tilde{r}_{iI_i}^{(n)})$. Also $OPT'_j(\bm{a}') -OPT_j(\bm{a}') \le \frac{2\delta'_{\min}}{5}$. As a result, we have $OPT_j(\bm{a}^{(\mu)}) -OPT_j(\bm{a}') =OPT_j(\bm{a}^{(\mu)}) -OPT'_j(\bm{a}^{(\mu)}) + OPT'_j(\bm{a}^{(\mu)})-OPT_j(\bm{a}') \overset{c}{\le} OPT_j(\bm{a}^{(\mu)}) -OPT'_j(\bm{a}^{(\mu)}) + OPT'_j(\bm{a}')-OPT_j(\bm{a}') \le \frac{4\delta'_{\min}}{5}$ in which $\overset{c}{\le}$ is in that $OPT'_j(\bm{a}')$ is optimal for edge server $j$ under the rewards $\{\Delta \tilde{r}_{ij}^{(n)}, i \in \mathcal{N}\}$. On the other side, $OPT_j(\bm{a}^{(\mu)}) -OPT_j(\bm{a}) \ge \delta'_{\min}, \exists j \in \mathcal{K}$. If  $OPT_j(\bm{a}^{(\mu)}) -OPT_j(\bm{a}') \le  \frac{4\delta'_{\min}}{5}, \forall j\in \mathcal{K}$, then it must be $OPT_j(\bm{a}^{(\mu)})= OPT_j(\bm{a}')$ for each server $j$, i.e., $\bm{I} = \bm{I}^{(\mu)}$. Therefore, $\bm{a}'$ is indeed the assignment $\bm{a}^{(\mu)}$ which is a $2$-approximate solution to  H-OAP of  Eq.~\eqref{Eq:opt_heter}.

Overall, the regret $\tilde{\mathcal{R}}(T)$ satisfies:
\begin{equation}
\begin{aligned}
\tilde{\mathcal{R}}(T) &\le \sum_{n=1}^{n_T}\Bigl(T_1\frac{N\overline{r}}{2}+K\frac{N\overline{r}}{2}+2NKe^{-n} \times 2^n \times \frac{N\overline{r}}{2}\Bigr)\\
& \le \Bigl(T_1\frac{N\overline{r}}{2}+K\frac{N\overline{r}}{2}\Bigr)n_T +\frac{2}{e-2}N^2K\overline{r}\\
& \le   \Bigl(T_1\frac{N\overline{r}}{2}+K\frac{N\overline{r}}{2}\Bigr)\log_2(T+2)+\frac{2}{e-2}N^2K\overline{r}\\
& \le  \Bigl(T_1\frac{N\overline{r}}{2}+K\frac{N\overline{r}}{2}\Bigr)\log_2(T+2)+4 N^2K\overline{r}\\
&=O(\log_2T),
\end{aligned}
\end{equation}
which completes the proof.
\end{appendices}
\end{document}